\documentclass[%
        10pt,
        twocolumn,
        nofootinbib,
 amsmath,amssymb,
 aps,
        prc,
        floatfix,
        tightenlines
]{revtex4-1}

\usepackage{graphicx}% Include figure files
\usepackage{dcolumn}% Align table columns on decimal point
\usepackage{bm}% bold math
\usepackage{bbm}
\usepackage{physics}
\usepackage{hyperref}% add hypertext capabilities
\usepackage[mathlines]{lineno}% Enable numbering of text and display math
\usepackage{color}

\hypersetup{breaklinks=true,colorlinks=true,linkcolor=blue,citecolor=blue,
filecolor=magenta,urlcolor=blue}

\makeatletter
\newcommand\newsubcommand[3]{\newcommand#1{#2\sc@sub{#3}}}
\def\sc@sub#1{\def\sc@thesub{#1}\@ifnextchar_{\sc@mergesubs}{_{\sc@thesub}}}
\def\sc@mergesubs_#1{_{\sc@thesub#1}}

\newcommand\newsupcommand[3]{\newcommand#1{#2\sc@sup{#3}}}
\def\sc@sup#1{\def\sc@thesup{#1}\@ifnextchar^{\sc@mergesups}{^{\sc@thesup}}}
\def\sc@mergesups^#1{^{\sc@thesup#1}}
\makeatother

\DeclareMathAlphabet{\mathbcal}{OMS}{cmsy}{b}{n}

\newcommand{\boldvec}[1]{\bm{#1}}

\newcommand{\ordervec}{\vec}

\newcommand{\inputvec}{\mathbf}

\newcommand{\xvec}{\boldvec{x}}
\newcommand{\kvec}{\boldvec{k}}

\newsubcommand{\ckvec}{\ordervec{c}}{k}

\newsubcommand{\bkvec}{\ordervec{b}}{k}
\newsubcommand{\ckvecset}{\ordervec{\inputvec{c}}}{k}

\newsubcommand{\ckvecapprox}{\mathbf{c}'}{k}
\newsubcommand{\ckvecapproxset}{\mathbf{C}'}{k}

\newsubcommand{\bkvecapprox}{\mathbf{b}'}{k}
\newsubcommand{\bkvecset}{\mathbf{B}}{k}
\newsubcommand{\bkvecapproxset}{\mathbf{B}'}{k}

\newcommand{\genobs}{y}

\newsubcommand{\genobsvec}{\ordervec{\genobs}}{k}
\newsubcommand{\genobsvecset}{\ordervec{\inputvec{\genobs}}}{k}

\newcommand{\LEC}{\lecs}

\newsubcommand{\akvec}{\mathbf{a}}{k}

\newsubcommand{\akvecapprox}{\mathbf{a}'}{k}
\newsubcommand{\akvecset}{\mathbf{A}}{k}
\newsubcommand{\akvecapproxset}{\mathbf{A}'}{k}

{}  % Remove the definition from the Physics package

\def\diffd{\mathrm{d}}  % Upright differentials
\DeclareDocumentCommand\differential{ o g d() }{ % Differential 'd'
    \IfNoValueTF{#2}{
        \IfNoValueTF{#3}
            {\diffd\IfNoValueTF{#1}{}{^{#1}}}
            {\mathinner{\diffd\IfNoValueTF{#1}{}{^{#1}}\argopen(#3\argclose)}}
        }
        {\mathinner{\diffd\IfNoValueTF{#1}{}{^{#1}}#2} \IfNoValueTF{#3}{}{(#3)}}
    }
\DeclareDocumentCommand\dd{}{\differential} % Shorthand for \differential

\newcommand{\pathd}{\mathcal{D}}  % differential symbol for path integrals

\DeclareDocumentCommand\pathdifferential{ o g d() }{ % Path 'D'
    \IfNoValueTF{#2}{
        \IfNoValueTF{#3}
            {\pathd\IfNoValueTF{#1}{}{^{#1}}}
            {\mathinner{\pathd\IfNoValueTF{#1}{}{^{#1}}\argopen(#3\argclose)}}
        }
        {\mathinner{\pathd\IfNoValueTF{#1}{}{^{#1}}#2} \IfNoValueTF{#3}{}{(#3)}}
    }

\newcommand{\kt}{\widetilde k}

\renewcommand{\LEC}{C_0}

\newcommand{\sss}{\scriptscriptstyle}
\newcommand{\VPTsss}{\sss\text{VPT}}
\newcommand{\DeltaVPT}{\Delta_{\VPTsss}}

\newcommand{\sigmaVPT}{\sigma_{\VPTsss}}

\newcommand{\beachball}{[\mathcal{BB}]}
\newcommand{\fatunity}{\mathbbm{1}}

\newcommand{\psiup}{\psi^{\phantom{\dagger}}_\uparrow}
\newcommand{\psidown}{\psi^{\phantom{\dagger}}_\downarrow}
\newcommand{\psiupD}{\psi^{\dagger}_\uparrow}
\newcommand{\psidownD}{\psi^{\dagger}_\downarrow}
\newcommand{\rhot}{\widetilde{\rho}}
\newcommand{\mut}{\widetilde{\mu}}

\newcommand{\kF}[0]{k_{\mathrm F}}

\begin{document}

\title{Testing Variational Perturbation Theory for Effective Actions\texorpdfstring{\\}{} Using the Gaudin-Yang Model}

\author{Pranav Sharma}
\email{sharma.1098@osu.edu}
\affiliation{Department of Physics, The Ohio State University, Columbus, OH 43210, USA}

\author{R.~J. Furnstahl}
\email{furnstahl.1@osu.edu}
\affiliation{Department of Physics, The Ohio State University, Columbus, OH 43210, USA}

\date{\today}
\begin{abstract}
  The background field formalism based on effective actions is a compelling framework for developing an effective field theory for nuclear density functional theory. 
  Among the challenges in carrying out this development is handling both the particle-hole and pairing channels beyond the mean-field level, which includes how to incorporate collective degrees of freedom.
  Here we use the exactly solvable one-dimensional Gaudin-Yang model as a theoretical laboratory to explore candidate approaches.
  We compare Variational Perturbation Theory (VPT) to ordinary many-body perturbation theory and the inversion method, 
  all to second order in their respective expansions,
  and verify issues with Hubbard-Stratonovich auxiliary fields.
  VPT outperforms the other approaches at this level over a wide range of densities. 
  The next steps to extend this approach toward nuclei are outlined.
\end{abstract}
\maketitle

\section{Introduction}\label{sec:introduction}

The phenomenological energy density functional (EDF) approach
has many successes in predicting ground-state properties and more across the table of nuclides~\cite{Bender:2003jk,Vretenar:2005zz,Colo:2020vik,Schunck:2019book}.
In the EDF or density functional theory (DFT) formulation%
\footnote{The differences between DFT as formalized for the Coulomb many-body problem and the nuclear EDF approach have been stressed by Duguet, Sadoudi, and others~\cite{Duguet:2010cv,Fraboulet:2022zvt}. We will not touch upon those issues in the current work and will use DFT and EDF interchangeably. }
of the nuclear many-body problem, the degrees of freedom (dofs) are low-resolution densities, currents, and collective fields,
which are 
appropriate for describing low-energy nuclear observables. 
Like other successful phenomenological approaches to nuclei, nuclear EDFs should contain the seeds of an effective field theory (EFT) formulation (see Table 1 in Ref.~\cite{Furnstahl:2019lue}). 
Various avenues leading to an EFT for DFT have been explored over the years (e.g., see Refs.~\cite{Drut:2009ce,Grasso:2016gls,Furnstahl:2019lue,Fraboulet:2022zvt} and references therein),
 but multiple challenges remain. 
In this paper and subsequent work we will advance a formulation using the background field formalism based on effective actions~\cite{Fukuda:1994pq,Valiev:1997bb,Drut:2009ce}.
Our particular focus here is on handling the particle-hole and pairing channels beyond the leading mean-field approximation.

Effective actions are the field theoretical framework for Legendre transforms~\cite{Zinnjustin:2002,Weinberg:1996II,Peskin:1995ev,Coleman:1985rnk,Srednicki:2007qs,Altland:2006,Stone:2000}, which in turn are the underlying basis for DFT~\cite{Lieb:1983,DREIZLER90,Parr:1994,Eschrig2003,engel2011density}.
In the background field approach, the quantum fields in the action are split into classical background fields and quantum fluctuations.
The effective action can be constructed as a functional of these background fields via Legendre transformations from sources coupled to individual quantum fields or composite (possibly nonlocal) operators; the ``mean fields'' or the various densities of nuclear EDFs can be understood to be the background fields arising as expectation values of these fields or composite operators.
This is a variational approach (the effective action is made stationary with respect to the background fields), which can be systematically improved. 
For example, if we add more sources
coupled to different operator densities, those sources will explore a richer variational space.
Casting nuclear DFT in an effective action framework naturally suggests generalizations and approximations, and sets the stage for an EFT treatment.
Past work by one of the authors and collaborators explored various aspects of this connection~\cite{Hammer:2000xg,Puglia:2002vk,Bhattacharyya:2004aw,Bhattacharyya:2004qm,Bhattacharyya:2006fg,Furnstahl:2002gt,Furnstahl:2006pa,Furnstahl:2004xn,Furnstahl:2007xm,Drut:2009ce}, which we will build on in the current work.

Modern phenomenological EDFs for nuclei are formulated as a realization of the Hartree-Fock-Bogoliubov (HFB) method, which is a self-consistent mean-field approach that treats both the particle-hole and pairing channels.
We take this as an essential feature of a nuclear EFT for DFT, so we seek to incorporate these correlations starting at leading order in our effective action expansion.
We also want to include beyond-mean-field physics~\cite{Grasso:2018pen}, which prominently includes the contribution of collective modes.
The standard tool for adding collective degrees of freedom to a quantum field theory formulation of a many-body system is the Hubbard-Stratonovich (HS) transformation~\cite{Fukuda:1994pq,Nagaosa:1999,Altland:2006,Furnstahl:2002gt}. 
The HS transformation immediately leads to a loop expansion of the effective potential, which is well suited for the sort of semi-classical approximation scheme that meshes with nuclear phenomenology. 
In principle we can accommodate collective modes in particle-hole and pairing channels by introducing multiple fields; in practice there are double counting issues~\cite{Nagaosa:1999,Altland:2006}.
Within an EFT in which modes can be cut off, this may be fixable~\cite{lan15,kug18,bon22}, but we have not yet explored such a formulation. 

The inversion method~\cite{Fukuda:1994pq,Fukuda:1995im} is another option for building a functional that incorporates HFB physics at leading order.
If sources are coupled to both the fermion density and pair density operators of the system, a 
double Legendre transformation with respect to the sources yields a functional of the c-number density and pair density.
This is DFT with pairing. 
This transformation can be carried out order by order in an EFT expansion parameter~\cite{Puglia:2002vk,Furnstahl:2006pa};
when applied to 
the grand canonical potential,
this is Kohn-Luttinger-Ward (KLW) inversion~\cite{FETTER71}.
Although it has no double-counting issues, this approach as applied so far is also perturbative in nature, as the self-consistency conditions are applied to standard perturbation theory, which may preclude efficiently including collective effects.

An alternative to these schemes is provided by Variational Perturbation Theory (VPT)~\cite{Kleinert_cqf:2018yjk,Kleinert_hst_paper:2011rb}.%
\footnote{There is more than one implementation of VPT in the literature.  We follow Kleinert here by introducing auxiliary fields as VPT parameters but discuss alternatives in Sec.~\ref{sec:summary}. See Refs.~\cite{Fraboulet:2021amf,Fraboulet:2022zvt} for other conceptually similar approaches to VPT.}
In this approach, \textit{classical} collective fields are introduced into the partition function and the field equations satisfied by these collective degrees of freedom are obtained order by order in a modified perturbation theory. 
Such an approach provides a functional of collective fields similar to the HS scheme, but in a manner that is agnostic towards channels and that is much simpler in higher orders because the fields are classical rather than quantum. 
The VPT expansion is also \textit{convergent} to the exact partition function at higher orders~\cite{Kleinert_hst_paper:2011rb}, in contrast to the asymptotic nature of standard perturbation theory.

Our strategy in the present work is to use the exactly solvable one-dimensional Gaudin-Yang
model~\cite{GAUDIN196755,PhysRevA.44.4915} as a theoretical laboratory to compare the candidate approaches numerically up to second order in each approach.
For clarity we work in the zero temperature thermodynamic limit, always keeping in mind the anticipated extension to functionals for finite systems and eventually nuclei in three dimensions. 
The paper is organized as follows: 
%In Sec.~\ref{sec:formalism} we review the general effective action formalism and the diagrammatic content of many-body perturbation theory, then present in turn the various approaches we will compare.
%In Sec.~\ref{sec:gaudin} we specialize the formalism to the Gaudin model, and  
In Sec.~\ref{sec:gaudin} we review the Gaudin-Yang model and present in turn the formalism for the various approaches we will compare as applied to the model.
In Sec.~\ref{sec:results} we present numerical results first for the exact, perturbative, and BCS energy density and chemical potential, followed by the corresponding results for the inversion method and VPT.
Our conclusions and plans for subsequent investigations are given in Sec.~\ref{sec:summary}.
Additional formal and numerical implementation details are given in two appendices.

\section{Gaudin-Yang model} \label{sec:gaudin}

\newcommand{\epstilde}{\tilde\epsilon}

\subsection{Exact results for the Gaudin-Yang model} \label{subsec:gaudin}

Our theoretical laboratory is a one-dimensional box of length $L$ with $N$ fermions of mass $m$ that interact via attractive pairwise contact forces, 
\begin{equation}
 H = -\frac{\hbar^2}{2m} \sum_{i=1}^{N} \frac{d^2}{dx_i^2} - \LEC \sum_{i<j = 1}^{N} \delta(x_i - x_j),
 \qquad \LEC > 0 .
\end{equation}
This is the same Hamiltonian as the leading-order EFT for a dilute system of fermions without spin-dependent interactions, which was studied for three dimensions in Refs.~\cite{Hammer:2000xg,Puglia:2002vk}.
Here, the restriction to leading order and one spatial dimension means there are no UV divergences in free space, so we do not introduce a regulator.
The number density is $\rho = N/L$ and after scaling $x$ by $\rho$ it is manifest that the ground-state properties depend only on the dimensionless coupling
\begin{equation}
    \lambda = \frac{m\LEC}{\hbar^2 \rho}\label{eq:gy_param} .
\end{equation}
Thus low density is strong coupling and high density is perturbative~\cite{PhysRevA.44.4915}.

We will assume $\nu=2$ distinct flavors.
The system can be solved using the Bethe ansatz~\cite{giamarchi03}; we simply quote the results. 
The exact dimensionless energy per particle from the Gaudin-Yang integral equations is~\cite{PhysRevA.44.4915} 
\begin{equation}
    \epstilde(\lambda) = -1 + \frac{4}{\pi} K^3 \lambda \int_{-1}^{1}dy\, y^2 f(y) ,\label{eq:casas_Epp}
\end{equation} 
where $f(y)$ and $K$ are found by simultaneously solving 
\begin{align}
    f(x) &= 2 - \frac{K}{\pi}
    \int_{-1}^{1}dy\, \frac{f(y)}{1 + K^2 (x-y)^2} ,
    \\
    \frac{1}{\lambda} &= \frac{K}{\pi}
    \int_{-1}^{1}dy\, f(y) . 
\end{align}
The perturbative expansion for the energy has been worked out to very high order in Ref.~\cite{Marino:2019fuy}. It is shown to second order alongside the exact solution and the BCS solution in
Sec.~\ref{sec:results}.

\subsection{Inversion method for the Gaudin model} \label{subsec:inv-gaudin}

The inversion method is an order-by-order Legendre transformation from the thermodynamic potential $W$ as a function of sources to the effective action $\Gamma$ as a function of the corresponding conjugate densities; see Appendix~\ref{sec:formalism} for details. %sources to the corresponding conjugate densities.
As discussed there, the effective action will correspond to the free energy functional in DFT.
Here, we outline the inversion procedure in the context of the Gaudin-Yang model.
%We outline the procedure and its connection to DFT here, in the context of the Gaudin-Yang model.
We rely heavily on the previous work applying the inversion method to dilute Fermi systems in traps~\cite{Bhattacharyya:2004aw,Bhattacharyya:2004qm,Puglia:2002vk}.

We begin with an action in the canonical ensemble %---again, 
with a contact interaction~\cite{Furnstahl:2006pa}: 
\begin{align}
    A = \int_x\psi^\dagger\qty[\partial_\tau - \frac{\nabla^2}{2m}]\psi - \LEC\int_x\psiupD\psidownD\psidown\psiup .
\end{align}
(The compact notation used here is defined at the start of Appendix \ref{sec:formalism}.)
To the Lagrangian, we add two source-density terms $j\bigl[\psiupD\psidownD + \psidown\psiup\bigr] - \mu\psi_\alpha^\dagger\psi_\alpha$, with the sources acting as constants since the ground state is time independent and uniform.
The effective action as a function of the density is obtained as a double Legendre transform in $\mu$ and $j$.
In the uniform system, the density and pair density are related to the particle number and pair condensate by factors of the spacetime volume ($\beta L$): 
\begin{align}
    \rho &= \big\langle\psi^\dagger\psi\big\rangle_{\mu, j} = -\frac{1}{\beta L}\pdv{W}{\mu} ,\\
    \phi &= \big\langle\psiupD\psidownD + \psidown\psiup\big\rangle_{\mu, j} = \frac{1}{\beta L}\pdv{W}{j} .
\end{align}
These are treated as zeroth order in the expansion and are thus not modified at higher order (see Eq.~\eqref{eq:inv_general_sc_eqs}).
The Legendre transformation is then 
\begin{align}
    \frac{1}{\beta L}\Gamma[\rho, \phi] = \frac{1}{\beta L}W[\mu, j] + \mu\rho -j\phi .
\end{align}
We get self consistency equations by demanding that the pairing source $j$ is zero in the true ground state, and by taking the density as zeroth order in correspondence with Kohn Sham DFT.
%The self-consistency equations that arise from the insertion of the expanded sources into $W$ % from Eq.~\eqref{eq:inv_general_sources} 
These self-consistency equations can be written in a simple manner because the system is uniform: 
\begin{align}
    \rho = \int\frac{\dd[2]{k}}{(2\pi)^2}\qty(1 - \frac{\xi_k}{E_k})\qc j_0 = -j_1 - j_2 - \ldots
\end{align}
where $\xi_k = \frac{k^2}{2m} - \mu_0$ and $E_k = \sqrt{\xi^2 + j_0^2}$.
The density is proportional to the particle number for a uniform system, so it is a conserved charge in the Gaudin-Yang model.
Since the density is of more immediate interest for the effective action calculation, we use the above form of the number equation, working in terms of the density instead of the particle number.

At zeroth order in the inversion, we compute $\Gamma_0 = W_0 + \mu_0\rho - j_0\phi$ with $W_0$ simply being the trace log of the quadratic part of the action (the inverse unperturbed Green's function). %unperturbed inverse Green's function.
Regulating the trace log by differentiating with respect to $\xi$ before doing the frequency integral and integrating back with respect to $\xi$ afterwards~\cite{Furnstahl:2002gt}, we get 
\begin{align}
            \frac{1}{\beta L}W_0 = \int\frac{\dd{k}}{2\pi}(\xi_k - E_k),
\end{align}

We then fix $\rho$ and $\phi$ from our expression for $W_0$: 
\begin{align}
    \rho &= \int\frac{\dd{k}}{2\pi}\qty(1 - \frac{\xi_k}{E_k})\\
    \phi &= -\int\frac{\dd{k}}{2\pi} \,\frac{j_0}{E_k} .
\end{align}
The zeroth order contribution to the energy density is then obtained by Legendre transformation. 
We rewrite $W_0$ to identify the effects of the Legendre transformation and identify pieces for comparison with other methods. 
\begin{align}
    \frac{1}{\beta L}W_0 &= \int\frac{\dd{k}}{2\pi} \frac{1}{E_k} \qty[\xi_k E_k - \xi^2 - j_0^2] \notag \\
    &= \int\frac{\dd{k}}{2\pi} \frac{1}{E_k} \biggl[\frac{k^2}{2m}\qty(E_k - \frac{k^2}{2m} + \mu_0)
    \notag \\
    & \qquad\qquad \null - \mu_0\qty(E - \mu_0 + \frac{k^2}{2m}) - j_0^2\biggr]   \notag\\
    &= \int\frac{\dd{k}}{2\pi}\qty[\frac{k^2}{2m}\qty(1 - \frac{\xi_k}{E_k})] - \mu_0\rho + j_0\phi
        ,
\end{align}

Thus, we find the lowest-order term taking the familiar form of kinetic energy multiplying an occupation function:
\begin{align}
    \frac{1}{\beta L}\Gamma_0 = W_0 + \mu_0\rho - j_0\phi = \int\frac{\dd{k}}{2\pi}\frac{k^2}{2m}\qty(1 - \frac{\xi_k}{E_k}) .
\end{align}
At first order, we get 
\begin{align}
    \frac{1}{\beta L}\Gamma_1 = -\frac{\LEC \rho^2}{4} - \frac{\LEC \phi^2}{4} ,
\end{align}
which allows us to immediately find the first-order contributions to the sources, 
\begin{align}
            \mu_1 &= \frac{1}{\beta L}\pdv{\Gamma_1}{\rho} = -\frac{\LEC}{2}\rho  , \\
            j_1 &= -\frac{1}{\beta L}\pdv{\Gamma_1}{\phi} = \frac{\LEC}{2}\phi .
\end{align}
Thus, we have 
\begin{align}
    \mu_1 &= -\frac{\LEC}{2}\int\frac{\dd{k}}{2\pi}\qty(1 - \frac{\xi_k}{E_k})  , \\
    j_0 &= -j_1 = \frac{\LEC}{2}\int\frac{\dd{k}}{2\pi} \frac{j_0}{E_k} .
\end{align}
These are the mean-field equations in the presence of pairing: the so-called ``number equation'' and the BCS gap equation.

To go to second order, we must calculate the beachball diagram. 
The anomalous diagram cancels against the Legendre transformation term (see  Fig.~\ref{fig:NLO_cancellation} in Appendix~\ref{subsec:inversion}), so we only need the beachball, whose contribution to $W_2$ we denote as $\overline{W}$.
The beachball diagram evaluates to

\begin{align}\label{eq:bb}
    &\frac{1}{2^3}\int\frac{\dd{k}}{2\pi}\frac{\dd{p}}{2\pi}\frac{\dd{q}}{2\pi} 
    \bigl(E_k + E_{k+q} + E_p + E_{p-q}\bigr)^{-1} \notag \\
    & \quad\null\times\bigl(uv_k uv_{k+q} uv_{p}uv_{p-q} - 2uv_{k+q}uv_{p-q}(v_k^2 u_p^2 + v_p^2 u_k^2) 
    \notag \\
    & \qquad\quad \null
    + u_k^2u_{k+q}^2v_p^2v_{p-q}^2 + v_k^2v_{k+q}^2u_{p}^2u_{p-q}^2\bigr) ,
    \end{align}
where the BCS-like occupation functions are defined for the purposes of this subsection as 
\begin{align}
    v_k^2 = \frac{1}{2}\qty(1 - \frac{\xi_k}{E_k})\qc u_k^2 = \frac{1}{2}\qty(1 + \frac{\xi_k}{E_k})\qc uv = -\frac{j_0}{2E_k}
.\end{align}

The beachball diagram is a function of $\mu_0$ and $j_0$ since it is calculated from the noninteracting propagator, but must be differentiated with respect to $\rho$ and $\phi$ (with the other held fixed) to get $\mu_2$ and $j_2$. 
For clarity, we demonstrate this procedure for calculating the derivatives of the sources with respect to $\phi$.
The derivatives with respect to $\rho$ follows similarly.
From the general thermodynamic relationships that define $\Gamma$, we have
\begin{align}
    \mu_2 &= \frac{1}{\beta L}\qty(\pdv{\Gamma_2}{\rho})_\phi \notag\\
    &= \qty(\pdv{\overline{W}_2}{\mu_0})_{j_0}\qty(\pdv{\mu_0}{\rho})_\phi + \qty(\pdv{\overline{W}_2}{j_0})_{\mu_0}\qty(\pdv{j_0}{\rho})_\phi\\
    j_2 &= -\frac{1}{\beta L}\qty(\pdv{\Gamma_2}{\phi})_\rho\notag\\
    &= \qty(\pdv{\overline{W}_2}{\mu_0})_{j_0}\qty(\pdv{\mu_0}{\phi})_\rho + \qty(\pdv{\overline{W}_2}{j_0})_{\mu_0}\qty(\pdv{j_0}{\phi})_\rho
\end{align}
To compute the derivatives of the sources with respect to the densities, we differentiate the number equation and the $\phi$ equation with respect to $\phi$ at constant $\rho$ to get
\begin{align}
    0 &= \int\frac{\dd{k}}{2\pi}\frac{j_0\xi_k}{E_k^3}\qty(\pdv{j_0}{\phi})_{\rho} + \int\frac{\dd{k}}{2\pi}\frac{j_0^2}{E_k^3}\qty(\pdv{\mu}{\phi})_{\rho}\\
    1 &= -\int\frac{\dd{k}}{2\pi}\frac{\xi_k^2}{E_k^3}\qty(\pdv{j_0}{\phi})_{\rho} - \int\frac{\dd{k}}{2\pi}\frac{j_0\xi_k}{E_k^3}\qty(\pdv{\mu_0}{\phi})_{\rho} .
\end{align}
These same integrals show up in the $\rho$ derivative, so we label them 
\begin{align}
    A = \int\frac{\dd{k}}{2\pi} \frac{j_0^2}{E_k^3}\qc
    B = \int\frac{\dd{k}}{2\pi} \frac{j_0\xi_k}{E_k^3}\qc
    C = \int\frac{\dd{k}}{2\pi} \frac{\xi_k^2}{E^3}\label{eq:inv_ABC}
    .
\end{align}
It is then a simple linear system of equations to solve for the derivatives we need, and we get 
\begin{align}
    \qty(\pdv{j_0}{\phi})_{\rho} &= \frac{A}{B^2 - CA}\\
    \qty(\pdv{\mu_0}{\phi})_{\rho} &= \frac{B}{CA - B^2} .
\end{align}
The chemical potential and pairing source can now be calculated via quadrature.
The second-order contributions to the energy, chemical potential, and ``gap'' $j_0$ are thus calculated from the beachball diagram as a function of $j_0$ and $\mu_0$, and the results are plotted against the results from the other approximation methods in Section~\ref{sec:results}.

\subsection{Hubbard-Stratonovich for the Gaudin-Yang model} \label{subsec:hst-gaudin}
The Hubbard-Stratonovich method integrates a quantum collective field into the partition function.
There are different channels that the collective field can be coupled to in this procedure, we apply it to the Gaudin-Yang model with a collective field in the particle-hole channel as this is the relevant channel to get the effective action dependent on the density.
For details, see Appendix~\ref{subsec:hst}.

In our 1D model, the partition function is
\begin{align}
    Z = \int\mathcal{D}\psi^\dagger\mathcal{D}\psi\, e^{-\int\dd[2]{x}\psi_\alpha^\dagger\qty[\partial_\tau - \frac{\nabla^2}{2}]\psi_\alpha + \LEC\int\dd[2]{x}\psi_{\uparrow}^\dagger\psi_{\downarrow}^\dagger\psi_{\downarrow}\psi_{\uparrow}} .
\end{align}
Multiplying the Boltmann factor by 1 in the form 
\begin{align}
    1 = \frac{1}{\mathcal{N}}\int\qty[\dd\sigma]e^{-\frac{1}{2}\int\dd[2]{x}\qty(\sigma - \LEC\psi^\dagger\psi)\frac{1}{\LEC}\qty(\sigma - \LEC\psi^\dagger\psi)},
\end{align}
where $\mathcal{N}$ is a pure Gaussian integral over $\sigma$,
the partition function becomes 
\begin{align}
    Z &= \int\mathcal{D}\psi^\dagger\mathcal{D}\psi\qty[\dd{\sigma}]e^{-\int\dd[2]{x}\psi_\alpha^\dagger\qty[\partial_\tau - \frac{\nabla^2}{2} - \sigma]\psi_\alpha - \int\sigma\frac{1}{2\LEC}\sigma}\\
    &= \int\qty[\dd{\sigma}]e^{\Tr\ln\qty(-\partial_\tau + \frac{\nabla^2}{2} + \sigma) - \int\sigma\frac{1}{2\LEC}\sigma}\notag\\
    &\equiv \int\qty[\dd{\sigma}]\exp{-A[\sigma]} .
\end{align}
We want to do a loop expansion in $\sigma$ about the saddle point of the action in the above path integral, so we need 
\begin{align}
    0 &= \fdv{\sigma}\qty[-\Tr\ln\qty(-\partial_\tau + \frac{\nabla^2}{2} + \sigma) + \int\sigma\frac{1}{2\LEC}\sigma]  \notag\\
    &= -\Tr\qty(-G\cdot \fdv{(-G^{-1})}{\sigma}) + \frac{1}{\LEC}\sigma\\
    &= -\rho + \frac{\sigma}{\LEC}\implies \sigma_0 = \LEC\rho,
\end{align}
where $\sigma_0$ denotes the field configuration at the saddlepoint. 

To perform a loop expansion, we separate our collective field into a classical part with quantum fluctuations $\sigma = \sigma_0 + \eta$ and expand in powers of $\eta$.
The lowest order term comes simply from the action itself evaluated at the saddlepoint: 
\begin{align}
    Z_{LO} = \exp{\Tr\ln\qty(-G_H^{-1}) - \beta L \frac{\sigma_0^2}{2\LEC}}.
\end{align}
The energy density is proportional to the negative log of the partition function, yielding
\begin{align}
    \mathcal{E}_{LO} &= -\frac{2}{\beta L} \int\dd[2]{x}\ln\qty{\qty(-\partial_\tau + \frac{\nabla^2}{2m} + \sigma_0)\delta_{x,x}} + \frac{\sigma_0^2}{2\LEC}  \notag\\
    &= -2\int\frac{\dd[2]{k}}{(2\pi)^2}\ln(ik^0 - \xi) + \frac{\sigma_0^2}{2\LEC}\qc \xi \equiv \frac{k^2}{2m} - \sigma_0.
\end{align}
Looking at the first term, we perform the $k^0$ integral first, regulating it by differentiating with respect to $\xi$ and then integrating back after the frequency integral is performed, just like in \cite{Furnstahl:2002gt}.
The result is 
\begin{align}
    N\int\frac{\dd{k}}{2\pi}\xi\theta(\kF - |k|) = \frac{2}{2\pi}\qty(\frac{\kF^3}{3} - 2\kF\sigma_0)
\end{align}
Recalling that in 1D without pairing, the Fermi momentum is related to the density by $\kF = \pi\rho/2$, the energy density is simplified to 
\begin{align}
    \frac{\pi^2\rho^3}{24} - \LEC\rho^2.
\end{align}
Including the $\sigma_0^2$ term, we ultimately end up with 
\begin{align}
    \mathcal{E}_{LO} = \frac{\pi^2\rho^3}{24} - \frac{\LEC\rho^2}{2}.
\end{align}

To go to NLO, we must integrate out the next terms in the $\eta$ expansion of $A[\sigma_0 + \eta]$. 
The linear terms vanish by construction,%
\footnote{We are expanding about a saddlepoint so the coefficient of the linear terms, the derivative of the action with respect to $\sigma$, must be zero.} 
so the next contribution comes from the quadratic $\sigma$ part and the second-order term in the expansion of the $\Tr\ln$:
\begin{align}
    Z_{NLO} &= e^{\Tr\ln(-G_H^{-1}) - \beta L\frac{\sigma_0}{\LEC}} \notag \\
    & \qquad\null\times
    \int\qty[\dd{\eta}]e^{-\int\eta\frac{1}{2\LEC}\eta - \frac{1}{2}\int G_H\eta G_H\eta}.
\end{align}
This is a bosonic gaussian in $\eta$, so we get a determinant to the power of $-1/2$. 
The NLO contribution is thus
\begin{align}
    \mathcal{E}_{NLO} - \mathcal{E}_{LO} = \frac{1}{2}\Tr\ln\qty(1 - 2\LEC\Pi)
.\end{align}
The Lindhard function $\Pi$ is calculated as in Ref.~\cite{Giuliani_Vignale_2005} and it is obtained in momentum space as 
\begin{align}
    \Pi(q^0, q) = \frac{-1}{2\pi \kF y}\ln(\frac{\nu^2 + \qty(\frac{1}{2}y + 1)^2}{\nu^2 + \qty(\frac{1}{2}y - 1)^2}),
\end{align}
Where $y \equiv \frac{q}{\kF}, \nu = \frac{q^0}{q \kF}$.
There is a logarithmic divergence at zero frequency for particles at the Fermi momentum here, and it leads to a divergence of a renormalon nature~\cite{Marino:2019wra} in the energy.
The NLO contribution of 
\begin{align}
    \frac{1}{2}\int \frac{\dd[2]{q}}{(2\pi)^2}\ln(1 + \frac{\LEC}{\pi \kF y}\ln(\frac{\nu^2 + \qty(\frac{1}{2}y + 1)^2}{\nu^2 + \qty(\frac{1}{2}y - 1)^2}))
\end{align}
has a series expansion in ring diagrams (powers of $\Pi$) that fail to converge in one dimension, despite the fact that the exact answer and perturbative approximations are finite.
Past the first term in the series (which corresponds to the Fock term), this NLO contribution is precisely the RPA energy. 
The failure of the RPA energy to converge is a unique feature of the one-dimensional Gaudin-Yang model and not a general feature of the Hubbard-Stratonovich program. We postpone details about multiple fields to future papers.

\subsection{VPT for a uniform system} \label{subsec:vpt-gaudin}
In contrast to the HST approach, the VPT framework~\cite{Kleinert_cqf:2018yjk, Kleinert_hst_paper:2011rb, Kleinert_piqm:2004ev, yukalov2021, Yukalov2019InterplayBA}, allows us to introduce fields for both the particle-hole channel ($\sigmaVPT$) \textit{and} the particle-particle channel ($\DeltaVPT$) in an \textit{exact} manner.
That is, we do not need to neglect (or suppress) regions of overlapping momentum integrations that would be overcounted amongst the different HS channels.
We accomplish this by adding and subtracting terms that couple our fermions to $\sigmaVPT$ and $\DeltaVPT$ as \textit{classical} fields, i.e., with no fluctuating quantum component.
A complete description of the workflow in VPT is given in Appendix~\ref{subsec:vpt}.  
For our application to the Gaudin-Yang model, we can write the action in a plane-wave basis and define 
\begin{equation}
  \xi \equiv \frac{k^2}{2m} - \mu - \sigmaVPT ,  
\end{equation} 
so we can write
\begin{align}
    Z &= \int\mathcal{D}\Psi^\dagger\mathcal{D}\Psi
    \exp{-A_0 - A_{\text{int}}}, %\notag \\
\end{align}
where our action in terms of Nambu spinors~\cite{Altland:2006} is 
\begin{align}
    A_0 &= \int
    \frac{\dd[2]{k}}{(2\pi)^{2}}
    \mqty[\psiupD & \psidown]
    \mqty[-i\omega +\xi & -\DeltaVPT\\ 
         -\DeltaVPT & -i\omega -\xi]
         \mqty[\psiup\\ \psidownD],\\
    A_{\text{int}} &= -\LEC \int
    \frac{\dd[2]{k}}{(2\pi)^{2}} \,
    \psiupD\psidownD
    \psidown\psiup
    \notag \\
    &\qquad+ \mqty[\psiupD & \psidownD]
    \mqty[\sigmaVPT & \DeltaVPT\\ \DeltaVPT & -\sigmaVPT]
    \mqty[\psiup \\ \psidownD] .\label{eq:vpt_gaudin_Aint}
    \end{align}
Expanding the definition of $\xi$ makes it manifest that the terms involving $\sigmaVPT$ and $\DeltaVPT$ cancel each other when writing down the full action $A_0 + A_{\text{int}}$.
However, if we perform perturbation theory with $A_0$ as the unperturbed action and $A_{\text{int}}$ as the perturbation, then to any finite order, the effects from $\sigmaVPT$ and $\DeltaVPT$ will \textit{not} cancel out for arbitrary configurations of these classical fields.
We thus perform perturbation theory with these modified actions and determine the values of $\sigmaVPT$ and $\DeltaVPT$ at each order of perturbation theory by applying the principle of minimal dependence to $W(\sigmaVPT, \DeltaVPT)$.
%As in the HS program, the advantage of introducing collective fields is that terms in the standard perturbation series are resummed based on the interactions of the collective degree of freedom instead of just counting powers of the coupling constant.

The modified propagator for the theory is 
\begin{align}
    G(x, y) = \int_k\frac{e^{ik\cdot(x - y)}}{(i\omega - E)(i\omega + E)}\mqty[-(i\omega + \xi) & \DeltaVPT\\ \DeltaVPT & -(i\omega - \xi)],\label{eq:vpt_gfs}
\end{align}
where $E = \sqrt{\xi^2 + \DeltaVPT^2}$.
The thermodynamic potential is calculated using the usual perturbative (cumulant) expansion, we just have extra vertices and a modified free propagator: 
\begin{align}
    Z &= Z_0\ev{e^{-A_{\text{int}}}}_0\\
      &= Z_0e^{-\ev{A_{\text{int}}}_0 + \frac{1}{2}\qty[\ev{A_{\text{int}}^2}_0 - \ev{A_{\text{int}}}_0^2] + \ldots}  \notag \\
    \implies W &= -\ln(Z) \notag \\
    &= -\ln(Z_0) + \ev{A_{\text{int}}}_0 \notag \\
    & \qquad \null - \frac{1}{2}\qty[\ev{A_{\text{int}}^2}_0 - \ev{A_{\text{int}}}_0^2]\label{eq:vpt_cumulant_exp} + \ldots
        .
\end{align}
All expectation values in this section will be with respect to $A_0$, so the subscript will be omitted from future brackets.

The zeroth-order term will come solely from the gaussian free action, which yields 
\begin{align}
    -\ln\det\mqty[i\omega - \xi & \DeltaVPT\\ \DeltaVPT & i\omega + \xi] = -\Tr\ln\mqty[i\omega - \xi & \DeltaVPT\\ \DeltaVPT & i\omega + \xi] .
\end{align}
The eigenvalues of the matrix are $i\omega \pm E$, so the trace becomes 
\begin{align}
    2\int\frac{\dd{k}}{(2\pi)} \ln(-\omega^2 - E^2)
        .
\end{align}
This can be regulated in the same manner as the inversion method's zeroth-order term, since they are mathematically the same expression.
The result is once again
\begin{align}
     -\Tr\ln\mqty[i\omega - \xi & \DeltaVPT\\ \DeltaVPT & i\omega + \xi] = \int\frac{\dd{k}}{2\pi}(\xi - E).
\end{align}
The difference here is just in the definition of $\xi$ since we now include collective fields instead of sources.
Although the fields act mathematically similar to sources at this stage, their contribution of additional vertices gives us a quantitatively different approximation scheme.
The differences will be seen explicitly at higher orders.
    
\begin{figure}[tbh]
    \centering
    \includegraphics[width=0.48\textwidth]{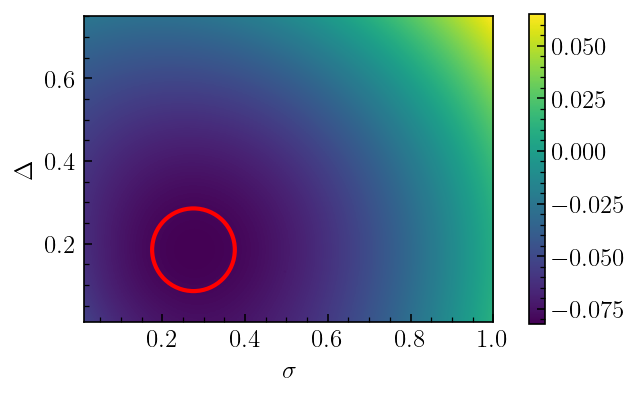}
    \includegraphics[width=0.48\textwidth]{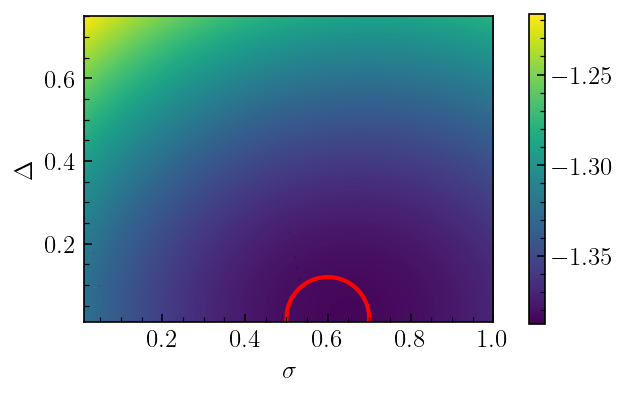}
    \caption{first-order approximation to $W = -ln(Z)$ as a function of $\sigmaVPT$ and $\DeltaVPT$, the auxiliary fields. The top figure yielded a density of $\rhot = 0.45$ and the bottom yielded a density of $\rhot = 1.2$. The minimum of the potential as a function of the VPT parameters can clearly be seen to change as a function of the chemical potential.}
        \label{fig:vpt_W1}
\end{figure}

The first-order term is the expectation value of the interaction term, Eq.~\eqref{eq:vpt_gaudin_Aint}.
Defining 
\begin{align}
    u_k^2 = \frac{1}{2}\qty(1 + \frac{\xi}{E})\qc v_k^2 = \frac{1}{2}\qty(1 - \frac{\xi}{E})\qc (uv)_k = \frac{\DeltaVPT}{2E}
        ,
\end{align}
it is a straightforward exercise in contour integration to show that the expectation value reduces to 
\begin{align}
    -\LEC\qty(\int\frac{\dd{k}}{2\pi}v_k^2)^2 - \LEC\qty(\int\frac{\dd{k}}{2\pi}(uv)_k)^2 + \notag\\2\sigmaVPT\int\frac{\dd{k}}{2\pi}v_k^2 + 2\DeltaVPT\int\frac{\dd{k}}{2\pi}(uv)_k
        .
\end{align}
It can also be seen from a partial fraction decomposition of the Green's functions in Eq.~\eqref{eq:vpt_gfs} that the two integrals in the above expression are proportional to the noninteracting%
\footnote{Unlike inversion, this is \emph{not} the proper expression for the density beyond the mean field level. There is no counting condition that keeps the density a zeroth order quantity here} particle ($\rhot$) and pair ($\phi$) densities respectively---they are equal to the Green's functions evaluated at the same point.
The exact relation is 
\begin{align}
    \rhot = 2\int\frac{\dd{k}}{2\pi}v_k^2\qc \phi = 2\int\frac{\dd{k}}{2\pi}(uv)_k
        .
\end{align}
The total first-order effective potential is then 
\begin{align}
    \Gamma_0 + \Gamma_1 &= \int\frac{\dd{k}}{2\pi}(\xi - E) + \mu\rhot - \frac{\LEC}{4}\qty(\rhot^2 + \phi^2) 
    \notag \\
    & \qquad\qquad \null + \sigmaVPT\rhot + \DeltaVPT\phi
        .
\end{align}
The thermodynamic potential $W$ can be plotted for various values of $\sigmaVPT$ and $\DeltaVPT$ as shown in Fig.~\ref{fig:vpt_W1}.
 
The principle of minimal dependence tells us that we must differentiate the result for $W_0 + W_1$ with respect to $\sigmaVPT$ and $\DeltaVPT$ and set them simultaneously to zero\footnote{if we cannot find a solution to these equations, then the minimal dependence comes from demanding the 2nd derivative is zero.}. 
This yields 
\begin{align}
    0 = &-\rhot 
    - \frac{\LEC}{2}\rhot\pdv{\rhot}{\sigmaVPT} 
    - \frac{\LEC}{2}\phi\pdv{\phi}{\sigmaVPT} \notag\\ 
    &+ \rhot 
    + \sigmaVPT\pdv{\rhot}{\sigmaVPT} 
    + \DeltaVPT\pdv{\phi}{\sigmaVPT}  \notag\\
    = &\qty(\sigmaVPT - \frac{\LEC}{2}\rhot)\pdv{\rhot}{\sigmaVPT} + \qty(\DeltaVPT - \frac{\LEC}{2}\phi)\pdv{\phi}{\sigmaVPT}\label{eq:1o_dsigma}\\
    0 = &-\phi 
    - \frac{\LEC}{2}\rhot\pdv{\rhot}{\DeltaVPT} 
    - \frac{\LEC}{2}\phi\pdv{\phi}{\DeltaVPT} \notag\\
    &+ \sigmaVPT\pdv{\rhot}{\DeltaVPT} 
    + \phi 
    + \DeltaVPT\pdv{\phi}{\DeltaVPT}  \notag\\
    = &\qty(\sigmaVPT - \frac{\LEC}{2}\rhot)\pdv{\rhot}{\DeltaVPT} + \qty(\DeltaVPT - \frac{\LEC}{2}\phi)\pdv{\phi}{\DeltaVPT}\label{eq:1o_dDelta}
    .
\end{align}
To have both of these equal to zero requires that $\sigmaVPT = \frac{1}{2}\LEC\rhot$ and $\DeltaVPT = \frac{1}{2}\LEC\phi$.
Note that the derivative of the zeroth-order term was canceled by the derivatives of the auxiliary field vertices at first order.
The energy per particle at this order is exactly what is obtained in the standard (HFB) mean field theory calculation when evaluated at the minimizing values of $\sigmaVPT$ and $\DeltaVPT$.
\begin{align}
    W_0+W_1 &= \int\frac{\dd{k}}{2\pi}\qty(\frac{\xi E - E^2}{E}) - \frac{\LEC\rhot^2}{4} - \frac{\LEC\phi^2}{4} 
        \notag \\ & \qquad \null
    + \sigmaVPT\rhot + \DeltaVPT\phi  \notag\\
    &= \int\frac{\dd{k}}{2\pi}\biggl[\frac{k^2}{2}\qty(1 - \frac{\xi}{E}) - (\mu + \sigmaVPT)\qty(1 - \frac{\xi}{E}) 
        \notag \\ & \qquad\qquad \null
    - \frac{\DeltaVPT^2}{E}\biggr] - \frac{\LEC\rhot^2}{4} - \frac{\LEC\phi^2}{4} 
        \notag \\ & \qquad\qquad \null
    + \sigmaVPT\rhot + \DeltaVPT\phi  \notag\\
    &= \frac{1}{4\pi}\int\!\dd{k}k^2\qty(1 - \frac{\xi}{E}) - \mu\rhot - \sigmaVPT\rhot - \DeltaVPT\phi
        \notag \\ & \qquad \null
     - \frac{\LEC\rhot^2}{4} + \frac{\LEC\phi^2}{4} + \sigmaVPT\rhot + \DeltaVPT\phi \notag\\ 
    &= \frac{1}{4\pi}\int\dd{k}k^2\qty(1 - \frac{\xi}{E}) - \mu\rhot - \frac{\LEC\rhot^2}{4} - \frac{\DeltaVPT^2}{\LEC} .
    \end{align}
Thus, the effective action is 
\begin{align}
    \Gamma_0 +\Gamma_1 &= W_1 + \mu\rho \notag \\
    & = \frac{1}{2\pi}\int\dd{k}\frac{k^2}{2m}\qty(1 - \frac{\xi}{E}) - \frac{\LEC\rho^2}{4} - \frac{\DeltaVPT^2}{\LEC} ,
\end{align}
meaning that the dimensionless energy per particle is 
\begin{align}
    \frac{8}{\LEC^2\rho}\Gamma = \frac{2}{\pi m\LEC^2\rho}\int\dd{k}k^2\qty(1 - \frac{\xi}{E}) - \frac{2\rho}{\LEC} - \frac{8\DeltaVPT^2}{\LEC^3\rho} .
\end{align}
This is exactly the BCS approximation result~\cite{Marino:2019wra}.

To go to second order, following Eq.~\eqref{eq:vpt_cumulant_exp}, we must now calculate 
\begin{align}
            \frac{1}{2}\qty[\ev{A_{\text{int}}^2} - \ev{A_{\text{int}}}^2] .
\end{align}
This is all connected vacuum diagrams with two vertices of either the auxiliary fields or the quartic coupling. 
All of the following calculations will involve ``mixed derivatives'': $\pdv{\rhot}{\DeltaVPT}$ and $\pdv{\phi}{\sigmaVPT}$. 
These work out to be the same term for the uniform system.
As such, we denote these mixed derivatives $M$.
The terms arising from expectation values with two auxiliary fields yield 
\begin{align}
    -\frac{\sigmaVPT^2}{2}\pdv{\rhot}{\sigmaVPT} - \frac{\DeltaVPT^2}{2}\pdv{\phi}{\DeltaVPT} - \sigmaVPT\DeltaVPT M .
\end{align}
The terms with one auxiliary field and one quartic coupling yield 
\begin{align}
    \frac{\LEC\DeltaVPT\phi}{2}\pdv{\phi}{\DeltaVPT} + \frac{\LEC\DeltaVPT\rhot}{2}M \notag\\+ \frac{\LEC\sigmaVPT\phi}{2}M + \frac{\LEC\sigmaVPT\rhot}{2}\pdv{\rhot}{\sigmaVPT} .
\end{align}
Finally, the terms with two quartic couplings yield 
\begin{align}
    -\frac{\LEC^2\rhot^2}{4}\pdv{\rhot}{\sigmaVPT} - \frac{\LEC^2\phi^2}{4}\pdv{\phi}{\DeltaVPT} - \frac{\LEC^2\rhot\phi}{2}M - \beachball .
\end{align}

Everything that is not the beachball $\beachball$ can be combined and simplified to a very compact form, yielding: 
\begin{align}
     -\frac{1}{2}\pdv{\rhot}{\sigmaVPT}\qty(\sigmaVPT - \frac{\LEC\rhot}{2})^2 - \frac{1}{2}\pdv{\phi}{\DeltaVPT}\qty(\DeltaVPT - \frac{\LEC\phi}{2})^2 \notag\\ 
     \null -M\qty(\sigmaVPT - \frac{\LEC\rhot}{2})\qty(\DeltaVPT - \frac{\LEC\phi}{2}) + \beachball .
\end{align}
These terms must all be differentiated with respect to $\sigmaVPT$ and $\DeltaVPT$ to find the (in principle, new) values of the auxiliary field parameters that minimize the dependence of $W$ on them.
The first-order action should in principle still be added to this to get the energy density, however for the sake of minimization, we know that the derivative of the first-order piece gives the number and gap equations in Eqs.~\eqref{eq:1o_dsigma}, \eqref{eq:1o_dDelta}. 
Furthermore, just as the derivatives of the first-order terms canceled the derivative of the zeroth order terms, the derivatives of some of the second-order terms cancel the derivatives of the first-order terms. 
In general, the terms that cancel the prior order's VPT equations come from the terms that involve VPT vertices.
Specifically, the differentiation of a VPT vertex at a given order will cancel a term from the prior order.
What is left is:
\begin{align}
    0 = &\frac{\LEC}{2}\Biggl\{\qty[\qty(\pdv{\rhot}{\sigmaVPT})^2 + \qty(\pdv{\phi}{\sigmaVPT})^2]\mathcal{N} \notag\\
    + &\pdv{\phi}{\sigmaVPT}\qty[\pdv{\rhot}{\sigmaVPT} + \pdv{\phi}{\DeltaVPT}]\mathcal{G}\Biggr\} \notag \\
    - &\frac{1}{2}\pdv[2]{\rhot}{\sigmaVPT}\mathcal{N}^2 
    - \frac{1}{2}\pdv{\phi}{\sigmaVPT}{\DeltaVPT}\mathcal{G}^2 \notag \\
    - &\pdv{\rhot}{\sigmaVPT}{\DeltaVPT}\mathcal{NG} + \pdv{\beachball}{\sigmaVPT}\label{eq:vpt2o_sigma}  \\
    0 = &\frac{\LEC}{2}\Biggl\{\pdv{\rhot}{\DeltaVPT}\qty[\qty(\pdv{\phi}{\DeltaVPT}) + \pdv{\rhot}{\sigmaVPT}]\mathcal{N} \notag\\
    + &\qty[\qty(\pdv{\phi}{\DeltaVPT})^2 
    + \qty(\pdv{\rhot}{\DeltaVPT})^2]\mathcal{G}\Biggr\}  \notag \\
    - &\frac{1}{2}\pdv{\rhot}{\DeltaVPT}{\sigmaVPT}\mathcal{N}^2 
    - \frac{1}{2}\pdv[2]{\phi}{\DeltaVPT}\mathcal{G}^2 \notag \\
    - &\pdv{\phi}{\DeltaVPT}{\sigmaVPT}\mathcal{NG} 
    + \pdv{\beachball}{\DeltaVPT}\label{eq:vpt2o_Delta} ,
\end{align}
        where the abbreviations $\mathcal{N} = \sigmaVPT - \frac{\LEC\rhot}{2}$, $\mathcal{G} = \DeltaVPT - \frac{\LEC\phi}{2}$ are made since they are the ``number'' and ``gap'' equations when set to 0, as they appear at the mean field level. 
        The beachball diagram is basically the same form as we had for inversion in Eq.~\eqref{eq:bb}, the only difference is $\xi$ is now defined as $\frac{k^2}{2m} - \mu - \sigmaVPT$
These must be solved at different $\mu$ values to find the second-order values of $\sigmaVPT, \DeltaVPT$.
These second-order values can then be inserted into the number equation to get a density $\rho$ from each $\mu$.

\section{Results} \label{sec:results}

% Fig.~\ref{fig:energy_full_range}, with an inset showing the difference between approximate and exact solutions.
% We follow conventions set by Refs.~\cite{PhysRevA.44.4915} and report results normalized by the two-particle binding energy (such that $\epstilde$ in Eq.~\eqref{eq:casas_Epp} is $\widetilde E/N$ when plotted as a function of $\rho$).
% Similar plots for the chemical potential are shown in Fig.~\ref{fig:mu_full_range}.

\begin{figure}[tbh]
        \centering
        \includegraphics[width=0.48\textwidth]{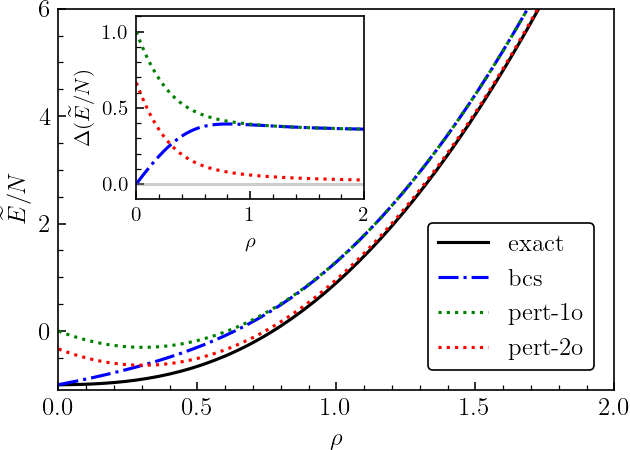}      
        \caption{The dimensionless energy per particle for the first- (green dots) and second-order (red dots) MBPT results compared to the exact (solid) and BCS (blue dot-dashed) solutions. Inset: Energies per particle  minus the exact solution.}
        \label{fig:energy_full_range}
\end{figure}

 \begin{figure}[htpb]
    \centering
    \includegraphics[width=0.48\textwidth]{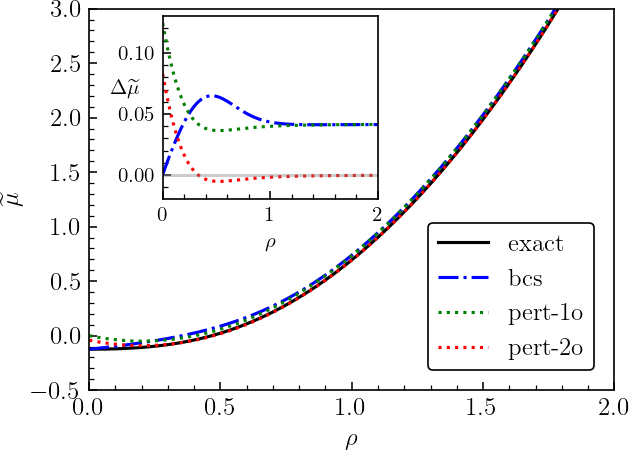}
    \caption{The dimensionless chemical potential, $\mut = \frac{\hbar^2}{m\LEC^2}\mu$, for first- (green dots) and second-order (red dots) MBPT compared to the exact (solid) and BCS (blue dot-dashed) solutions. Inset: Chemical potential  minus the exact solution.}
    \label{fig:mu_full_range}
\end{figure}

In this section we compare results for the various methods from Sec.~\ref{sec:gaudin} to the exact results. 
Following the conventions of Ref.~\cite{PhysRevA.44.4915}, we will plot the energy per particle (or energy density) in dimensionless units by scaling it by the exact energy per particle at zero density.%
\footnote{The binding energy of a pair is $B = \frac{-m\LEC^2}{4\hbar^2}$, and the exact energy at zero density is $\frac{N}{\nu}B$. In our analysis, the degeneracy $\nu = 2$. We use units where $\hbar = m = 1$, and plot at $\LEC=1$, which is still general because of Eq.~\eqref{eq:gy_param}.}
In Fig.~\ref{fig:energy_full_range} we plot the  results for the energy from previous work~\cite{PhysRevA.44.4915,Quick_BCS1D.47.11512}, with the inset showing the difference between approximate and exact solutions. 
The corresponding plots for the chemical potential are shown in Fig.~\ref{fig:mu_full_range}. 
The inclusion of pairing (BCS) at leading order leads to a deviation from first-order perturbation theory for densities less than one (in dimensionless units). 
The BCS energy agrees with the strong coupling limit at zero density, but deviates immediately.

Recent work examining resurgence has taken the perturbative expansion for the Gaudin-Yang model to very high order~\cite{Marino:2019fuy,Marino:2019wra}. 
For our ultimate application to nuclear DFT, high-order perturbation theory is unlikely to be feasible, so we focus only on results up to second order.

     \begin{figure}[tbh]
        \centering
        \includegraphics[width=0.48\textwidth]{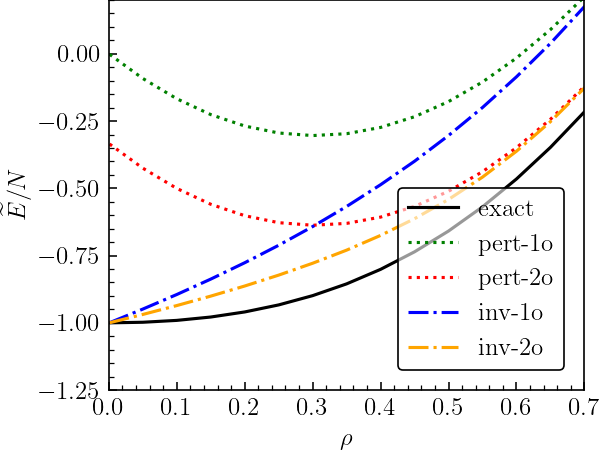}
        \caption{The strong coupling regime of the dimensionless energy per particle as a function of dimensionless density (i.e., $C_0=1$) for the  first- and second-order MBPT compared to the first- and second-order inversion results, as well as the exact solution.}
        \label{fig:energy_inv_compare}
    \end{figure}

\begin{figure}[tbh]
    \centering
    \includegraphics[width=0.48\textwidth]{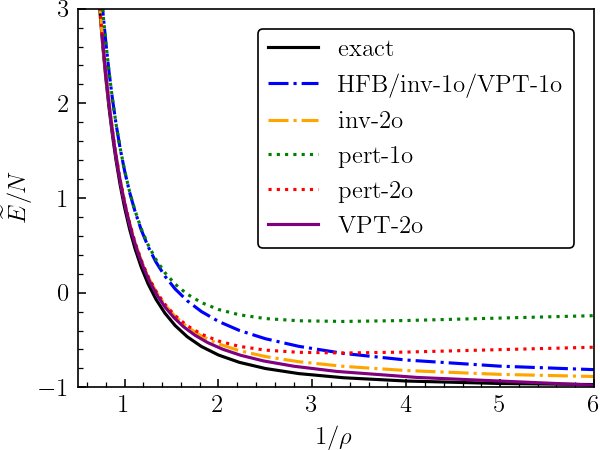}
    \includegraphics[width=0.48\textwidth]{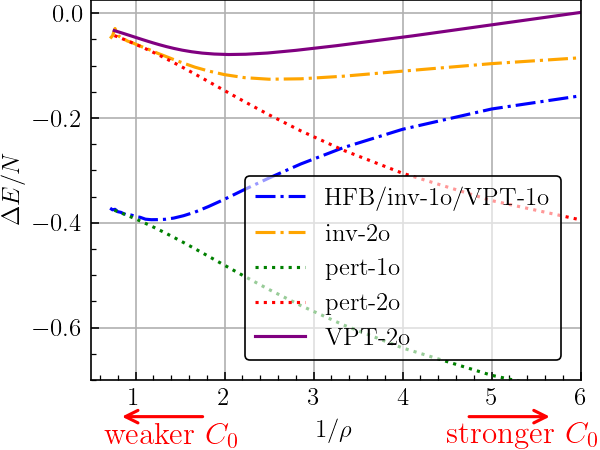}
    %\label{fig:delta_Epp}
    \caption{(a) The energy per particle for the first- and second-order inversion method results compared against the exact and BCS solutions and VPT second order. (b) Same as (a) but minus the exact. Both are plotted against the dimensionless inverse density.}
    \label{fig:inv_energy_density}
\end{figure}

The dimensionless energy per particle for the inversion method at first and second order are shown at low density in Fig.~\ref{fig:energy_inv_compare} and over a larger range of inverse density in Fig.~\ref{fig:inv_energy_density}.
The inversion method is a systematic way to do perturbation theory in both the particle-hole and pairing channels.
As noted in Sec.~\ref{subsec:inv-gaudin}, this leads to the agreement of the first-order inversion and BCS results if we make the association that $j_0$ is the BCS gap.
Because of the inclusion of pairing, at second order the inversion method equals or outperforms ordinary second-order MBPT (the beachball diagram) at all densities and is substantially closer to the exact results than at first order.
This sets a standard for second-order results for comparison to VPT.

The dimensionless energies per particle for VPT at first and second order are shown in Figs.~\ref{fig:inv_energy_density} and 
compared to perturbation theory and the inversion method.
We encountered significant numerical issues at low densities, in the strong coupling region, and so only show results to $\rho \approx 0.17$.
As noted in Sec.~\ref{subsec:vpt-gaudin}, the first-order VPT results coincide with BCS.
The second-order VPT energy is substantially closer to the exact results at all densities that have been calculated.
The improvement over second-order inversion reflects the resummation of perturbation theory induced by the field equations of the classical collective VPT fields in both channels.

\begin{figure}[tbh]
    \centering
    \includegraphics[width=0.48\textwidth]{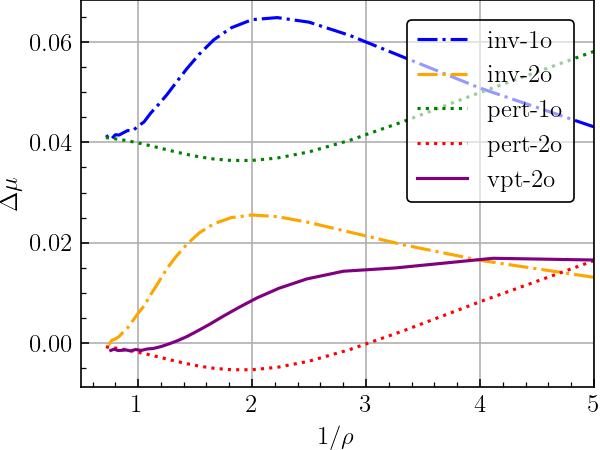}
    \caption{Difference between exact and approximate dimensionless chemical potentials versus the inverse density, calculated through second order in inversion, perturbation theory, and VPT.}
    \label{fig:mu_vs_density}
\end{figure}

The differences between exact and approximate dimensionless chemical potentials versus the inverse density are shown in Fig.~\ref{fig:mu_vs_density} for the same approximations as in Fig.~\ref{fig:inv_energy_density}(b).
All of these calculations manifest thermodynamic consistency.
This occurs in inversion by construction, see Eq.~\eqref{eq:inv_sources_from_Gamma}.
To calculate the VPT effective action in a thermodynamically consistent manner, some care must be taken to invert from dependence on $\mu$ to dependence on $\rho$ when Legendre transforming.
The way we inverted the dependencies (as explicitly discussed in Appendix \ref{subsec:vpt}) is thermodynamically consistent.

\begin{figure}[tbh]
    \centering
    \includegraphics[width=0.48\textwidth]{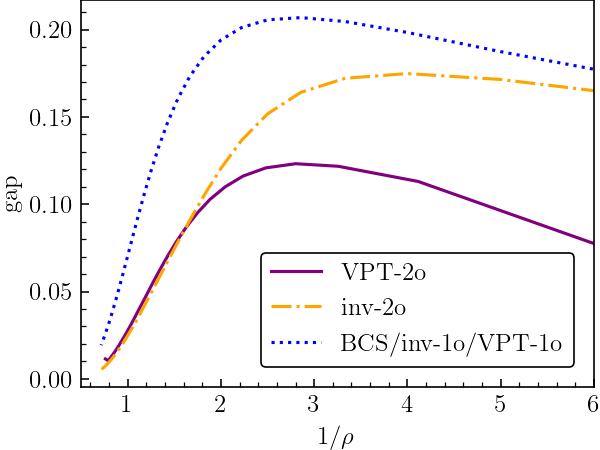}
    \caption{Gaps in the single-particle energy spectrum at second order in inversion and VPT.}
    \label{fig:gaps_vs_density}
\end{figure}

In Fig.~\ref{fig:gaps_vs_density}, we compare ``gaps'' from the various methods.
The gaps are $j_0 = -j_1 - j_2$ in the inversion method and $\Delta$ in the VPT method.
Second-order values for both of these were obtained, and we find that in both methods, the gap is suppressed when beyond-mean-field effects are included.
The beachball diagram is manifestly\footnote{The beachball is the sole contribution to $\Gamma_2$ in the inversion scheme so $j_2$ is only from the beachball, and the VPT equations at second order are seen in Eqs.~\eqref{eq:vpt2o_sigma} and \eqref{eq:vpt2o_Delta} to be solved by the number and gap equations if the beachball diagram was 0} responsible for deviations from the BCS gaps in both methods, this can be seen to arise since the beachball diagram adds coupling between elements of the generalized density matrix~\cite{Blaizot:1985}, 
\begin{align}
    \mqty(\big\langle\psiupD(x)\psiup(y)\big\rangle & \big\langle\psiupD(x)\psidownD(y)\big\rangle\\ \big\langle\psidown(x)\psiup(x)\big\rangle & \big\langle\psidown(x)\psidownD(x)\big\rangle) ,
\end{align}
which did not occur at first order in each scheme. 

The suppression of the pairing gap to second order in the inversion scheme is known in the three-dimensional case~\cite{Furnstahl:2006pa} as the Gor'kov--Melik-Barkhudarov correction~\cite{Gorkov:1961, Heiselberg2000}. 
Here, we do not get a scaling constant due to differences in phase space yielding a qualitatively and quantitatively different contribution to the pairing and perturbative regimes of the model.
In the case of VPT to second order, we find that the gap is much smaller than the inversion result in the perturbative limit, dropping by a factor of nearly two at the furthest computed point of the strong coupling limit.

\section{Summary and outlook} \label{sec:summary}

In this paper, we explore different methods of constructing effective actions in the background field formalism, including contributions systematically from both the particle-hole and pairing channels, 
using the one-dimensional Gaudin-Yang model as a testbed. 
Our analysis considered MBPT, the inversion method, the loop expansion with an HS field, and the VPT expansion. 
In all three cases that were testable for Gaudin-Yang (see Sec.~\ref{subsec:hst-gaudin}), the energy per particle as a function of the density improved in going from first order (mean field) to second order in the respective expansions. 
As clearly demonstrated in Fig.~\ref{fig:inv_energy_density}, the second-order VPT expansion performed the best. 
The systematic incorporation of both channels with collective fields that couple to the fermions is quite promising as an avenue for arriving at a nuclear EDF with beyond-mean-field physics in multiple channels. 

The VPT prescription and the inversion prescriptions are both quite flexible, as they can accommodate couplings to any desired composite field. 
Furthermore, unlike the naive HS method,
both of these methods naturally avoid the ambiguity of channel selection.
With VPT, it is also easy to see how to handle many-body forces (i.e., three-body forces, four-body, etc.) with collective fields.

Our overarching
path forward toward nuclear DFT is to next consider finite but trapped one-dimensional systems to demonstrate how to implement the prescriptions when the densities (and thus the collective fields) have space dependence.
From there, we plan to generalize to three-dimensional uniform and finite trapped systems, and then finally reach realistic nuclear systems. 
There are multiple avenues to take in this direction.
The VPT prescription can be used to find an optimal Green's function~\cite{You2003} instead of providing collective fields.
We also plan to adapt the inversion method for a VPT calculation to align the VPT method more closely with the formulation of Kohn-Sham DFT. 
Finally, we hope to examine and extend work on integrating in multiple Hubbard-Stratonovich fields with cutoffs on momentum integration. 
There exists some work~\cite{lan15, kug18, bon22} that takes this approach for condensed matter systems, using the functional Renormalization Group with the cutoffs.

There remain further challenges for the formulation of EDFs as an EFT.
The question of power counting is not directly
addressed via any of the methods considered here, though the exponentially convergent nature of the VPT expansion may be of significance in determining a power counting parameter. 
Projection of symmetry broken states is also not addressed by the inclusion of collective fields. 
As progress is made toward power counting and symmetry restoration schemes, they should be integrated with the collective mode formalisms assessed here.

\begin{acknowledgments}
We thank P.G.\ Reinhard, S.~Sundberg, H. Merritt, M. Hisham, and S. Jaiswal for engaging discussions.  This research was supported by the National Science Foundation Award Nos.\ PHY-2209442 and PHY-2514765 and by the NUCLEI SciDAC program under award DE-FG02-96ER40963. 
RJF also acknowledges support from the ExtreMe Matter Institute EMMI
at the GSI Helmholtzzentrum für Schwerionenforschung GmbH, Darmstadt, Germany.\end{acknowledgments}

\clearpage

\appendix

\section{Formalism} \label{sec:formalism}

\subsection{Background field formalism for effective actions}

Throughout this work, we use a four-vector notation for Euclidean spacetime $x \equiv \{\tau,\xvec\}$%
, with the $\tau$ dependence usually implicit, or $\kt \equiv \{k_0,\kvec\}$
in momentum space.
The formalism in this section is presented in $D$ Euclidean spacetime dimensions for generality.
For the sake of compactness, we frequently abbreviate integration measures and functional dependencies as 
\begin{align}
    \int\dd[D]{x}f(x, y)g(x) &\to \int_x f_{x, y} g_x ,
    \\ 
    \int\frac{\dd[D]{k}}{(2\pi)^D}G(k)&\to \int_k G_k .
\end{align}
Functional matrix multiplication is also occasionally abbreviated with dot product notation: 
\begin{align}
    \int_{x_1} f(x_1)g(x_1) &\to f\cdot g\qc \notag\\\int_{y_1, y_2} A_{x_1}B_{x_1,x_2}C_{x_2} &\to A\cdot B\cdot C .
\end{align}

To formulate an energy density functional from a quantum field theory~\cite{Fukuda:1994pq,Valiev:1997bb,Puglia:2002vk}, we start with the partition function for the theory with external sources coupled to composite densities, which defines a corresponding generating functional, and then Legendre transform from a functional of sources to a functional of densities.
This is only one possible effective action that can be computed from the theory as we can add sources for any type of field (composite or not) and then invert the source dependence to arrive at another effective action.
For conventional DFT,  we must at least Legendre transform with respect to the density, though other sources can be added and inverted, as illustrated for pairing in Sec~\ref{subsec:inversion}.

Explicitly, the partition function for a quantum field theory with action $A$ 
is of the form (normalization factors are implicit)
\begin{align}
    Z[J(x)] \equiv e^{-\beta W[J(x)]} = \int\mathcal{D}[\psi^\dagger]\mathcal{D}[\psi]e^{-A + J(x)\cdot\psi^\dagger\psi},
    \label{eq:PI_general}
\end{align}
where we include the minimally required source coupled to the density operator. 
We use the Matsubara formalism at inverse temperature $\beta$, but we will only consider the zero temperature ($\beta\rightarrow\infty$) limit.  

The action will be of the form 
\begin{align}
    A = \int\dd[D]{x}\psi_\alpha^\dagger(x)\qty(\pdv{\tau} - \frac{\nabla^2}{2m} + v_{\text{ext}})\psi_\alpha(x) + A_{\text{int}}
    ,
\end{align}
with external potential $v_{\text{ext}}(x)$ and interaction action $A_{\text{int}}$.
The thermodynamic potential $W[J]$ is extracted from the logarithm of the partition function, and we can use the convexity~\cite{Zinnjustin:2002} of the exponential to write 
\begin{align}
    \ln\ev{e^{-J\cdot\psi^\dagger\psi}} \geq J\cdot\ev{\psi^\dagger\psi}  .
\end{align}
Denoting the expectation value of $\psi^\dagger\psi$ as $\rho$, the left side can be rewritten as 
\begin{align}
    -W[0] + W[J] \geq -J\cdot\rho\notag \\
    \implies W[0] \leq W[J] + J\cdot\rho  .
\end{align}
We define the (functional) Legendre transform of $W[J]$ as 
\begin{align}
    \Gamma[\rho] = W[J] + \int_x J_x\rho_x
    \label{eq:legendre_transform},
\end{align}
and note that
\begin{align}
    W[0] \leq \Gamma[\rho] .
    \label{eq:W_inequality}
\end{align}
In the zero temperature limit ($\beta\rightarrow\infty$), the partition function will be dominated by the ground state (assumed to be non-degenerate), so $W[0]$ becomes the ground state energy of the system $E_0$
and the inequality \eqref{eq:W_inequality} becomes~\cite{Zinnjustin:2002} 
\begin{align}
    E_0 \leq \Gamma[\rho].
\end{align}
Thus, $\Gamma[\rho]$ is a 
functional of the density, whose minimum is the ground-state energy, i.e., it is an EDF.

In the notation of Ref.~\cite{engel2011density}, the thermodynamic potential $W[J]$ from Eq.~\eqref{eq:legendre_transform} acts like the energy functional with external potential $v_\text{ext}(x)$
\begin{align}
    E[v_\text{ext}[n]] = F[n] + \int_x v_\text{ext}(x)\rho(x)\label{eq:hk-notation},
\end{align}
and its Legendre transformation with respect to the density $n(x)$ acts as the HK functional $F[n]$.
The correspondence between Eq.~\eqref{eq:legendre_transform} and Eq.~\eqref{eq:hk-notation}
is manifest, and the HK theorems thus apply to the choice of effective action employed here.
The source $J$ couples to the density operator in the same way as the external potential in the action, but $J$'s role is as a fictitious source that is taken to zero at the end of the calculation.

The key step that makes this choice of effective action work as an EDF is the inversion of the relationship between $\rho$ and $J$, since 
\begin{align}
    \fdv{W}{J} &= -\ev{\psi^\dagger\psi} = -\rho,\\
    \fdv{\Gamma}{\rho} &= \int_x\fdv{W}{J}\fdv{J}{\rho} + \int_x\fdv{J}{\rho}\rho + \int_xJ\fdv{\rho}{\rho} = J.
\end{align}
This relationship makes explicit that taking $J$ to zero is 
equivalent to demanding that the effective action be stationary.
The fulfillment
of these derivative relations is entirely dependent on the ability for $J$ to be expressed as a functional of $\rho$,%
\footnote{This is guaranteed by the strict concavity of $W[J]$~\cite{Valiev:1997bb}.}  
which is the key point of both the construction of the effective action and the construction of the energy density functional.
Indeed, the invertibility of the functional relationship is precisely the first HK theorem~\cite{engel2011density}!

The second Hohenburg-Kohn theorem is also automatically included: the universality of the functional is ensured naturally, since the external potential can be trivially separated from the microscopic theory as it couples to the density in the same way as the source; the remaining interacting quantum field theory is the same regardless of the form of $v_\text{ext}$.
Explicitly, by reading off the definition in Eq.~\eqref{eq:PI_general} we see that 
\begin{align}
    W[v_{\text{ext}} = 0, J] &= W[v_{\text{ext}}, J - v_{\text{ext}}]\\
    \implies \Gamma_{v_\text{ext}=0}[\rho] &= W[0, J] + \int_x J(x)\rho(x) \notag \\
    & \qquad \null - \int_x v_{\text{ext}}(x)\rho(x)
.\end{align}
The external potential can be separated from the effective action in exactly the same way as the source, the rest of the functional is explicitly unchanged from its definition in Eq.~\eqref{eq:PI_general}. 
The goal of calculating an energy density functional is thus the same as trying to calculate the effective action from a theory of interacting fermions. 

\begin{figure*}[tbh]
    \centering
    \includegraphics[width=0.55\textwidth]{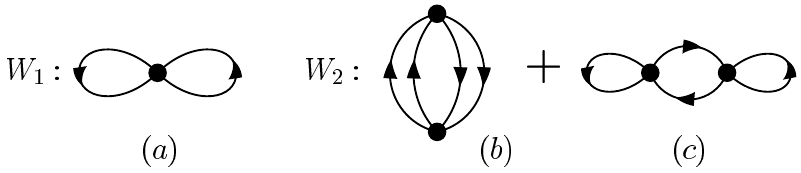}
    \caption{Hugenholtz diagrams for the unrenormalized (a) $W_1$ and (b), (c) $W_2$ functionals in a homogeneous, dilute Fermi system with an attractive delta function interaction. The vertices are all given by $-C_0$. As defined here, positive $C_0$ corresponds to attraction.}
    \label{fig:hugenholtz_diagrams}
\end{figure*}

A difficulty~\cite{Nagaosa:1999,Kleinert_hst_paper:2011rb, Altland:2006} in carrying out this program is the presence of competing channels of collective correlations. 
In particular, we have correlations in both the particle-hole and pairing channels.
At the mean-field level, we have Hartree-Fock-Bogoliubov theory as a way of incorporating pairing correlations~\cite{ringschuck}, but going beyond mean field with both channels has not been systematically explored in the effective action framework.

The simplest approximation scheme for the calculation of the effective action is using many-body perturbation theory (MBPT). 
In the path integral framework, we define MBPT as a Taylor expansion in the interacting part of the action~\cite{Altland:2006}: 
\begin{align}
    Z &= \int\mathcal{D}\psi^\dagger\mathcal{D}\psi\,e^{-(A_0 + A_1)} \notag \\
    &= \int\mathcal{D}\psi^\dagger\mathcal{D}\psi\,e^{-A_0}\qty(1 - A_1 + \frac{1}{2}A_1^2 + \ldots) \notag \\
    &= Z_0\ev{1 - A_1 + \frac{1}{2}A_1^2\ldots}_0,
\end{align}
where the expectation value is with respect to $Z_0$, defined by the action $A_0$.
Taking the (negative) log of the partition function yields the thermodynamic potential $W$ in the grand canonical ensemble, or the free energy $F$ in the canonical ensemble, up to factors of $\beta$.
For a uniform system, we can also extract a factor of the space volume $\mathcal{V}$ in however many spatial dimensions the system occupies to get an energy density.
Regardless of the ensemble used, the log of the partition function can be expanded in powers of $A_1$, which manifests the linked cluster theorem (see, e.g., Ref.~\cite{Negele:1988vy}) in the path integral framework.
To second order, we have 
\begin{align}
   -\ln(Z) = -\ln(Z_0) + \ev{A_1} - \frac{1}{2}\qty[\ev{A_1^2} - \ev{A_1}^2],
\end{align}
which can be Legendre transformed with respect to the chemical potential in the grand canonical ensemble from the grand canonical potential to an effective action.
In the canonical ensemble, the Legendre transformation is implicit, though pairing effects are not addressed at the mean-field level in this case.
Regardless, the choice of the ensemble does not change the MBPT expansion, it only affects what is done with the resultant log expansion.

\subsection{Inversion method}\label{subsec:inversion}

Here we outline the order-by-order inversion procedure and its connection to DFT.
%The inversion method is an order-by-order Legendre transformation from sources to the corresponding conjugate densities.
% We outline the procedure and its connection to DFT here, then specialize to the Gaudin-Yang model in Sec.~\ref{subsec:inv-gaudin}.
% We rely heavily on the previous work applying the inversion method to dilute Fermi systems in traps~\cite{Bhattacharyya:2004aw,Bhattacharyya:2004qm,Puglia:2002vk}.
%
We begin with an action in $D$ spacetime dimensions for fermions in the canonical ensemble.
For concreteness and connections to effective field theory potentials, we use a contact interaction, but the method applies for general potentials.
We thus have:
\begin{align}
    A = \int_x \psi_\alpha^\dagger\qty[\partial_\tau - \frac{\nabla^2}{2m} - v(x)]\psi_\alpha - {C_0}\int_x \psi_{\uparrow}^\dagger\psi_{\downarrow}^\dagger\psi_{\downarrow}\psi_{\uparrow}
\end{align}
To this, we add spacetime-dependent sources coupled to the density and pair density operators\footnote{Without a magnetic field, it is sufficient to take a real valued $j$ that couples to both $\psi^\dagger\psi^\dagger$ and its conjugate. Generalizing to separate sources works the same way, but it is not needed to demonstrate the procedure.}: 
\begin{align}
    A \to A - \int_x\mu_x\psi_{\alpha,x}^\dagger\psi_{\alpha,x} + \int_xj_x\qty(\psi_{\uparrow,x}^\dagger\psi_{\downarrow,x}^\dagger + \psi_{\downarrow,x}\psi_{\uparrow,x}).
\end{align}
The density source acts directly as a local chemical potential. 
While the standard thing to do is include a chemical potential in addition to the source, incorporating the chemical potential into the source is convenient for the application to infinite matter in Sec.~\ref{subsec:inv-gaudin}.
This will make our partition function --- and thus our thermodynamic potential --- a functional of $\mu(x)$ and $j(x)$. 
In particular, the thermodynamic potential is defined as $W[\mu, j] = -\ln(Z[\mu, j])$ so that 
\begin{align}\label{eq:inv_general_dW}
    \fdv{W[\mu, j]}{\mu(x)} = -\big\langle\psi_\alpha^\dagger\psi_\alpha\big\rangle\qc \fdv{W[\mu,j]}{j(x)} = \big\langle\psiupD\psidownD + \psidown\psiup\big\rangle
\end{align}
We thus need to perform a double Legendre transform so we have 
\begin{align}
    \Gamma[\rho_x, \phi_x] = W[\mu_x, j_x] + \int_x\mu_x\rho_x - \int_xj_x\phi_x,
    \label{eq:double_LT}
\end{align}
where the density is $\rho(x) = \big\langle\psi_\alpha^\dagger\psi_\alpha\big\rangle$ and we denote the pair density $\phi(x) = \big\langle\psiupD\psidownD + \psidown\psiup\big\rangle$.

\begin{figure*}[tbh]
    \centering
    \includegraphics[width=0.7\textwidth]{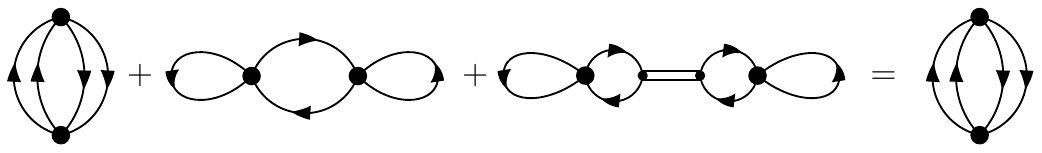}
    \caption{Cancellation of the anomalous diagram by the Legendre transformation term at NLO. The double lines represents the inverse of $[\partial^2W_0/\partial\mu^2]_{\mu=\mu_0}$. See Ref.~\cite{Furnstahl:2006pa} for details.}
    \label{fig:NLO_cancellation}
\end{figure*}

We assert that the sources and potentials have expansions in 
an EFT parameter.
The expansion is not strictly a power series, e.g., logarithmic corrections are allowed as is usual in an EFT expansion. 
Expanding in orders of the parameter, we have 
\begin{align}
    W[\mu, j] &= W_0[\mu, j] + W_1[\mu, j] + W_2[\mu, j]+\ldots\\
    \mu &= \mu_0 + \mu_1 + \mu_2 + \ldots\\
    j &= j_0 + j_1 + j_2 + \ldots\\
    \Gamma[\rho, \phi]&= \Gamma_0[\rho, \phi] + \Gamma_1[\rho, \phi] + \Gamma_2[\rho, \phi] + \ldots.
\end{align}
We carry out inversion by substituting these expansions in Eqs.~\eqref{eq:inv_general_dW} and \eqref{eq:double_LT} and equating equal orders, treating $\rho$ and $\phi$ as order unity.
In the general case, at zeroth order, we get
\begin{align}
    \Gamma_0[\rho, \phi] = W_0[\mu_0, j_0] + \mu_0\cdot\rho - j_0\cdot\phi.
\end{align}
This implies the following conditions on the zeroth order sources~\cite{Drut:2009ce}:
\begin{align}
    \rho(x) = -\qty(\fdv{W_0[\mu_0,j_0]}{\mu_0(x)})_{j_0}\qc
    \phi(x) = \qty(\fdv{W_0[\mu_0,j_0]}{j_0(x)})_{\mu_0} ,\label{eq:inv_general_sc_eqs}
\end{align}
with these densities unchanged at higher order, as expected in Kohn-Sham DFT.
The $W_0$ term is the sum of eigenvalues of a system that has no inter-particle interactions, but has the potentials modified by the zeroth order sources.

In the true ground state $j(x) = 0$, which imposes a relationship between the terms in the expansion of the pairing source: 
\begin{align}
    j_0(x) = -j_1(x) - j_2(x) - \ldots,
    \label{eq:inv_general_sources}
\end{align}
with a similar condition on the non-chemical potential part of $\mu(x)$.
These equations for $j$ and $\mu$ (through the definition of $\rho$) become self-consistency conditions for the DFT.

We obtain the $n$th-order approximation to the effective action by inserting the $n$th-order expansions for the sources into the $n$th-order perturbative approximation to $W$ (diagrammatically represented through second order by Fig.~\ref{fig:hugenholtz_diagrams}) and Taylor expanding to get all terms of order $n$, adding them to the terms of order $n$ from the Legendre transformation.
So to go to first order, 
\begin{align}
    \Gamma_1 = W_1 + \mu_1\cdot\fdv{W_0}{\mu_x}\eval_{\mu_0} + j_1\cdot\fdv{W_0}{j_x}\eval_{j_0} + \mu_1\cdot\rho - j_1\cdot\phi\label{eq:inv_1o_cancellation},
.\end{align}

The first-order part of $W$ is going to be 
\begin{align}
    W_1 &= \LEC\ev{\int_x \psiupD\psidownD\psidown\psiup} \notag\\
    &= \int_x\qty[-\big\langle\psi^{\phantom{\dagger}}_\uparrow\psiupD\big\rangle\big\langle\psidownD\psidown\big\rangle + \big\langle\psiup\psidown\big\rangle\big\langle\psidownD\psiupD\big\rangle] \notag \\
    &= \frac{\LEC}{4}\int_x\qty[\rho^2 + \phi^2] ,
\end{align}
where spacetime dependence is notationally suppressed and the substitutions for the density and pair density can be made since we have an unpolarized system.
The first-order piece can be conveniently expressed directly in terms of the density and pair density already, so the inversion is done at this order. 

The Taylor-expanded part of $W_0$ - the second and third terms in Eq.~\eqref{eq:inv_1o_cancellation} - is going to cancel with the Legendre transform terms according to Eq.~\eqref{eq:inv_general_sc_eqs}.
This happens at all orders, and is demonstrated diagrammatically at second order in Fig.~\ref{fig:NLO_cancellation}.
What is left is 
\begin{align}
    \Gamma_1[\rho(x), \phi(x)] = \frac{\LEC}{4}\int_x\qty[\rho^2 + \phi^2] .
\end{align}
To calculate the next order of the source expansions, we can take derivatives of the effective action since by definition of the Legendre transform, 
\begin{align}
    \mu_i(x) = \fdv{\Gamma_i[\rho,\phi]}{\rho(x)}\qc j_i(x) = -\fdv{\Gamma_i[\rho, \phi]}{\phi(x)}\label{eq:inv_sources_from_Gamma}.
\end{align}
Applying this to our first-order effective action, we get 
\begin{align}
    \mu_1(x) &= \frac{\LEC}{2}\rho(x)\qc \\
    j_1(x) &= -\frac{\LEC}{2}\phi(x)\implies j_0(x)=\frac{\LEC}{2}\phi(x) .
\end{align}
Proceeding to the next order follows this same reasoning. 
The explicit dependence on $\mu_i, j_i$ vanishes through the Legendre transform as indicated above for $i>1$.
The reflection of the higher-order effects directly manifests itself through $j_0$ and $\mu_0$ having more complicated self-consistent equations to satisfy (in the uniform case, $\mu_0$ is determined by a simple number equation).

The zeroth-order sources that appear in $W_0$ act as local Kohn-Sham potentials~\cite{Oliveira:1988}: they are sources that reflect the interacting system, but they only occur at the zeroth order (non-interacting) level.
The inversion method thus has the attractive feature of a direct correspondence between path integral methods and existing DFT techniques.

\subsection{Hubbard-Stratonovich Transformation}\label{subsec:hst}

The basic idea of the HS transformation is to introduce a quantum collective field into the partition function by exploiting the translational invariance of the (bosonic) functional gaussian integral to decouple the two-body interaction $V(x, y)$:
\begin{widetext}
\begin{align}
    \int\mathcal{D}\sigma \exp{-\int_{x,y} \sigma(x)V^{-1}(x, y)\sigma(y)} &=\int\mathcal{D}\sigma\exp{-\int_{x,y}\qty(\sigma_x - i\int_z\psi_{\alpha, z}^\dagger\psi_{\alpha, z} V_{x, z})_x V_{x,y}^{-1}\qty(\sigma - i\int_z\psi_{\beta, z}^\dagger\psi_{\beta, z} V_{z, y})}\nonumber\\
    \implies \fatunity &= \frac{\displaystyle\int\mathcal{D}\sigma\, \exp{-\int_{x,y}(\sigma - iV\psi_\alpha^\dagger\psi_\alpha)_x V_{x,y}^{-1}(\sigma - iV\psi_\beta^\dagger\psi_\beta)_y}}{\displaystyle\int\mathcal{D}\sigma\, \exp{-\int_{x,y} \sigma(x)V^{-1}(x, y)\sigma(y)}}\label{eq:hst-unity}
    .\end{align}
\end{widetext}
In particular, we take 
\begin{align}
    Z = \int\mathcal{D}\psi^\dagger\mathcal{D}\psi\, e^{-A} \to  \int\mathcal{D}\psi^\dagger\mathcal{D}\psi\, e^{-A}\cdot \fatunity ,
\end{align}
with $\fatunity$ written as in Eq.~\eqref{eq:hst-unity}.
The denominator of this ``fat unity''~\cite{Altland:2006} is simply a normalization factor and can be absorbed into the normalization for the path integral. 
The exponents in the numerator combine so we have 
\begin{align}
            A &\to \int_x\psi^\dagger_{\alpha, x} G_0^{-1}(x, y)\psi_{\alpha,x} 
            -2i\int_{x}\psi_{\alpha,x}^\dagger\psi_{\alpha, x}\sigma_x 
            \notag \\ 
            & \qquad
            + \int_{x,y}\sigma_xV^{-1}(x, y)\sigma_y\label{eq:hst_action} .
\end{align}
The equation of motion for $\sigma$ can be read off as a functional derivative of Eq.~\eqref{eq:hst_action}, which tells us that at the classical level, 
\begin{align}
    2i \int_x\psi_{\alpha, x}^\dagger\psi_{\alpha, x}V(x, y) = \sigma_y  .
\end{align}
This is a density operator convoluted with the potential,\footnote{The factor of $i$ is not included for a strictly negative potential.} it would be a pair density operator being convoluted if we decoupled in the particle-particle channel.
The action in the path integral is a functional of the new collective field, and we can expand around the classical minimum of the action $A$, integrating out quantum fluctuations in $\sigma$. 
Doing so order-by-order gets us a better approximation to our partition function and thus to our effective action.
The strategy is summarized as follows.
\begin{enumerate}
    \item Find the minimum of $A[\sigma]$.
    The minimizing field configuration $\sigma_0$ will be expressible in terms of the density.
    If we were to use a field in the pairing channel, we would instead find that the minimum is expressible in terms of the pair density $\phi\sim\ev{\psi^\dagger_\uparrow\psi^\dagger_\downarrow}$.
    \item Write the collective field as $\sigma_0(x) + \eta(x)$ such that $\ev{\eta} = 0$; the functional integral is now over $\eta$.
    \item Approximate the partition function by integrating out $\eta$ contributions to a given order. 
    At lowest order, we disregard the $\eta$ integral entirely and the partition function is approximated by the saddle point.
    At the next order, we treat the theory as a free field theory in $\eta$, integrating out the quadratic part of the remaining action.
    At higher orders, we include higher powers of $\eta$ as vertices, using the quadratic part as a propagator.
\end{enumerate}

We now illustrate the procedure through NLO for a non-uniform system.
The fermions can be integrated out exactly because the auxiliary field decouples the quartic part of the action.
The resultant Grassmann determinant is still a functional of $\sigma$ and is thus re-exponentiated as a $\Tr\ln$ so we have a partition function for the $\sigma$ field:
\begin{align}
    Z = \int\mathcal{D}\sigma \exp{\Tr\ln G^{-1} - \int_{x, y}\sigma_x V_{x, y}^{-1}\sigma_y} ,
\end{align}
where $G^{-1}_{x, y} = {G_0^{-1}}_{x, y} - 2i\sigma_x\delta_{x, y}$.
We want a semi-classical approximation so we look for the saddle point of the resultant action which is just a functional of $\sigma$ now: 
\begin{align}
    \fdv{\sigma(z)}\qty{-\Tr\ln G^{-1} + \int_{x, y}\sigma_xV_{x, y}^{-1}\sigma_y} &= 0\\
    \implies 2\nu iG(z, z) + 2\int\dd{y}V^{-1}(z, y)\sigma(y) &= 0 .
\end{align}
The factor $\nu$ is the degeneracy of the fermions, for a spin-$\frac{1}{2}$ system $\nu$ is 2. 
Specializing to the case of an attractive contact interaction $V(x, y) = -\LEC\delta(x - y)$, the saddle point equation becomes 
\begin{align}
    \nu iG(z, z) = \frac{1}{\LEC}\sigma_0(z),
\end{align}
where the subscript denotes this as the field configuration that minimizes the classical action.
This is directly proportional to the density.
So the auxiliary field at the saddle point is equal to the density scaled by the coupling.
Here, the mean-field equations arise from the claim that the macroscopic field configuration should, neglecting quantum fluctuations, extremize the classical action describing the microscopics of the theory.
We now split the collective field $\sigma$ into its expectation value in the presence of a source, plus fluctuations: $\sigma = \sigma_0 + \eta$ with $\ev{\sigma} = \sigma_0$.
The trace log can then be expanded as 
\begin{align}
    &\Tr\ln(G_0^{-1} - 2i\sigma_0(x) - 2i\eta(x))
    \notag  \\
    & \qquad\equiv \Tr\ln(G_H^{-1} - 2i\eta(x)) \notag \\
    & \qquad= \Tr\ln(G_H^{-1}) + \Tr\ln(1 - 2iG_H\cdot\eta) \notag \\
    & \qquad= \Tr\ln(G_H^{-1}) - \Tr\Bigl[\sum_n \frac{1}{n}(2iG_H\cdot\eta)^n\Bigr],
\end{align}
where $G_H$ denotes the Hartree propagator, implicitly defined by the first and second lines. 
The path integral is then over the fluctuations 
\begin{align}
    Z &= Z_0\int\mathcal{D}\eta\, e^{-\eta\cdot\qty[V^{-1} - G_H(x,y)G_H(y,x)]\cdot\eta 
    + \mathcal{O}(\eta^3)},
\end{align} 
where the quadratic part has been explicitly written for the sake of obtaining the propagator of the collective fluctuations, and 
\begin{align}
    Z_0 = e^{-\Tr\ln{G_H^{-1}(x, y) + \int_{x, y}\sigma_0(x) V_{x, y}^{-1}\sigma_0(y)}}\label{eq:hs_ph_Deta} .
\end{align}
The lowest-order (LO) contribution to the effective action will then be only the action evaluated at the saddlepoint with no fluctuations:
\begin{align}
    \Gamma_{LO} = \qty[\Tr\ln{G_H^{-1}} - \int\sigma V^{-1}\sigma]_{\sigma = \sigma_0},
\end{align}
with the next-to-lowest-order (NLO) contribution being the ``free'' part of the new collective field theory defined by the path integral in Eq.~\ref{eq:hs_ph_Deta}: 
\begin{align}
    \Gamma_{NLO} &= \Gamma_{LO} - \frac{1}{2}\ln\det[D^{-1}(x, y)]  \notag \\
    &= \Gamma_{LO} - \frac{1}{2}\Tr\ln[D^{-1}(x, y)], 
\end{align}
where 
\begin{equation} 
  D^{-1}(x,y) = V^{-1}(x,y) + G_H(x,y)G_H(y,x).
\end{equation}
Going to higher orders would be doing perturbation theory in the fluctuations using this new propagator. 
The NLO contribution can be recognized as the Fock term and the RPA sum by expanding the log in a series~\cite{Furnstahl:2002gt}. 
Furthermore, there is a spin sum involved in the LO term that does not appear in the NLO term as the NLO term arises from a single component scalar propagator being integrated.
As such, the LO terms have a power of $\nu$ that the NLO term will not. 
Proceeding to higher orders shows that successive orders are organized by powers of $\nu$, in a ``large-$N$ expansion'', as discussed in Ref.~\cite{Furnstahl:2002gt}.

The HS field is collective and directly relates to the density, which is an appealing feature for the discussion of DFT from a path integral perspective.
However, because the HS field is a quantum field, we must integrate over phase space to include the effects of fluctuations. 
This presents a difficulty when we have two active channels instead of one: we must integrate over fluctuations in both fields and the domains in phase space are not strictly disjoint, particularly at high momenta, naively leading to double counting~\cite{Nagaosa:1999,Altland:2006}.

\subsection{Variational Perturbation Theory}\label{subsec:vpt}

In contrast to the HST approach, the VPT framework~\cite{Kleinert_cqf:2018yjk, Kleinert_hst_paper:2011rb, Kleinert_piqm:2004ev, yukalov2021, Yukalov2019InterplayBA}, allows us to introduce fields for both the particle-hole channel ($\sigmaVPT$) \textit{and} the particle-particle channel ($\DeltaVPT$) in an \textit{exact} manner.
That is, we do not need to neglect (or suppress) regions of overlapping momentum integrations that would be overcounted amongst the different HS channels.
We accomplish this by adding and subtracting terms that couple our fermions to $\sigmaVPT$ and $\DeltaVPT$ as \textit{classical} fields, i.e., with no fluctuating quantum component.  
As in the case of the inversion method, we combine the chemical potential and the source in anticipation of the testbed system we analyze in Sec.~\ref{sec:gaudin}. 
In a finite system, these will be separate terms.
Writing our action with Nambu spinors~\cite{Altland:2006} 
we define
\begin{widetext}
\begin{align}
 A_0 &\equiv \int_x\mqty[\psiupD & \psidown]\mqty[\partial_\tau - \frac{\nabla^2}{2m} - \mu(x) - \sigmaVPT(x) + v(x) & -\DeltaVPT(x)\\ -\DeltaVPT(x) & \partial_\tau + \frac{\nabla^2}{2m} + \mu(x) + \sigmaVPT(x) - v(x)]\mqty[\psiup\\ \psidownD]\label{eq:vpt_gen_free}\\
 A_{\text{int}} &\equiv - \LEC\int_x\psiupD\psidownD\psidown\psiup + \int_x\DeltaVPT\qty[\psiupD\psidownD + \psidown\psiup] + \sigmaVPT\qty[\psiupD\psiup - \psidown\psidownD] .
\end{align}
\end{widetext}
The terms involving $\sigmaVPT$ and $\DeltaVPT$ can be seen to trivially cancel each other when writing down the full action $A_0 + A_{\text{int}}$.
However, if we perform perturbation theory with $A_0$ as the unperturbed action and $A_{\text{int}}$ as the perturbation, then to any finite order, the effects from $\sigmaVPT$ and $\DeltaVPT$ will \textit{not} cancel out for arbitrary configurations of these classical fields.
As in the HS program, the advantage of introducing collective fields is that terms in the standard perturbation series are resummed based on the interactions of the collective degree of freedom instead of just counting powers of the coupling constant.

The idea that drives the reorganization of perturbation theory in this scheme is that the original problem does not depend on $\sigmaVPT$ and $\DeltaVPT$, so we minimize the dependence of our effective potential on $\sigmaVPT$ and $\DeltaVPT$ at each order of perturbation theory in our calculation of the effective action. %
Originally, this was only done to first order and found to be variational, so it was named Variational Perturbation Theory. Going to higher orders, it is found that minimizing dependence on the auxiliary parameters is not necessarily variational anymore. The minimizing should be thought of as a \emph{principle of least dependence}~\cite{Kleinert_piqm:2004ev}, not a variational principle, contrary to the name.
The principle of least dependence is what drives the approximation scheme here: to best approximate the true effective action, we want the thermodynamic potential it is Legendre transformed from (i.e. $W$) to depend as minimally as possible on the auxiliary fields.`
The motivation here is that the true thermodynamic potential with all orders of perturbation theory summed up does not depend on the auxiliary fields, so to best approximate the true potential with the truncated sum at a finite order of perturbation theory, we should minimize the dependence on the parameters.

The general workflow for the VPT expansion at a given order is: 
\begin{enumerate}
    \item Calculate the thermodynamic potential $W = -\ln(Z)$ to the given order with the modified propagator and additional vertices. The result will be a functional of $\sigmaVPT$ and $\DeltaVPT$ as well as $\mu$. 
    \item Minimize the dependence of $W$ on $\sigmaVPT$ and $\DeltaVPT$ for a given configuration of $\mu$. Ideally, this is by finding $\sigmaVPT$ and $\DeltaVPT$ configurations that satisfy 
    \begin{equation}
        \qty(\fdv{W[\mu,\sigma,\Delta]}{\sigmaVPT(x)})_{\mu, \Delta} = 0\qc \qty(\fdv{W[\mu,\sigma,\Delta]}{\DeltaVPT(x)})_{\mu, \sigma} = 0 .
    \end{equation} 
    (In some cases, one must look for solutions where the second derivatives are zero instead; for an example, see Sec~5.15 of Ref.~\cite{Kleinert_piqm:2004ev}.) 
    \item Calculate the density of the system with the optimal field configurations $\sigma^*$ and $\Delta^*$
        \begin{align}
            \rho(x) = -\fdv{W[\mu, \sigma^*[\mu], \Delta^*[\mu]]}{\mu(x)}
        \end{align}
        and use it to Legendre transform the thermodynamic potential to the effective action: 
        \begin{align}
            \Gamma[\rho] = W[\mu[\rho(x)]] + \int_x \mu_x\rho_x
        .\end{align}
        For the uniform system discussed in this paper, inverting to get dependence on $\rho$ is simple to do parametrically. 
        In the more general non-uniform system, the details of such a procedure need to be explored. 
        A scheme adapting the inversion method from Sec.~\ref{subsec:inversion} is one way to proceed.
    \end{enumerate}

    As an illustration of this program in a more general setting, we 
    fill in the details to first order.
    The zeroth order part of $W$ is given by the trace log of the free action in Eq.~\eqref{eq:vpt_gen_free}: 
    \begin{widetext}

    \begin{align}
        W_0[\mu, \sigma, \Delta] = -\Tr\ln(-\mqty[\partial_\tau - \frac{\nabla^2}{2m} - \mu(x) - \sigmaVPT(x) + v(x) & -\DeltaVPT(x)\\ -\DeltaVPT(x) & \partial_\tau + \frac{\nabla^2}{2m} + \mu(x) + \sigmaVPT(x) - v(x)])
    \end{align}
    And the first-order term is given by $\ev{A_\text{int}}$:
    \begin{align}
        W_1[\mu, \sigma, \Delta] = -\LEC&\int_x\big\langle\psiupD\psidownD\psidown\psiup\big\rangle
        + \int_x\DeltaVPT\big\langle\psiupD\psidownD 
        + \psidown\psiup\big\rangle
        + \sigmaVPT\big\langle\psiupD\psiup - \psidown\psidownD\big\rangle.
    \end{align}
    \end{widetext}
    Due to the inclusion of the classical collective fields in the free action, we get contributions from the original (quartic) interaction term in both channels in a similar manner to the inversion method, but we now get extra interaction terms arising from the coupling of the fermions with the collective fields.
    Denoting the Nambu Green's function as $G_{\alpha\beta}$ for the sake of compactness, we have
    \begin{align}
        \ev{A_\text{int}} = -\LEC&\int_x\qty[G_{12}G_{21} - G_{11}G_{22}] -\notag \\
        &\int_x\qty[\DeltaVPT(G_{12} + G_{21}) 
        + \sigmaVPT(G_{11} - G_{22})].
    \end{align}
    The VPT prescription now is to minimize the dependence of $W_0[\mu, \sigma, \Delta] + W_1[\mu, \sigma, \Delta] = -\Tr\ln(-G^{-1}) + \ev{A_\text{int}}$ on $\sigmaVPT$ and $\DeltaVPT$ by setting the functional derivatives to 0. 
    Doing so yields 
    \begin{align}
       0 =& -\fdv{G_{21}}{\sigmaVPT}\qty(\DeltaVPT + \LEC G_{12}) - \fdv{G_{12}}{\sigmaVPT}\qty(\DeltaVPT + \LEC G_{21}) \notag\\
        &+ \fdv{G_{11}}{\sigmaVPT}\qty(\sigmaVPT - gG_{22}) + \fdv{G_{22}}{\sigmaVPT}\qty(\sigmaVPT + \LEC G_{11})\\
        0 = &-\fdv{G_{21}}{\DeltaVPT}\qty(\DeltaVPT + \LEC G_{12}) - \fdv{G_{12}}{\DeltaVPT}\qty(\DeltaVPT + \LEC G_{21}) \notag\\
        &+ \fdv{G_{11}}{\DeltaVPT}\qty(\sigmaVPT - \LEC G_{22}) + \fdv{G_{22}}{\DeltaVPT}\qty(\sigmaVPT + \LEC G_{11})
    \end{align}
    For an unpolarized system, $G_{11}(x, x) = -\frac{\rho(x)}{2} = -G_{22}(x, x)$ and $G_{ij}(x,x) = -\frac{\phi(x)}{2}$ for $i\neq j$.
    As such, to simultaneously solve both of the VPT conditions above, we must have 
    \begin{align}
        \DeltaVPT(x) = \frac{\LEC}{2}\phi(x)\qc \sigmaVPT = \frac{\LEC}{2}\rho(x),
    \end{align}
    which are the BCS equations.
    Denoting the field configurations that solve these equations as $\sigma'$ and $\Delta'$, we use these to calculate the density
    \begin{align}
        \rho(x) = \sum_\alpha\ev{\psi_\alpha^\dagger\psi_\alpha} = -G_{11}^{\sigma', \Delta'}(x, x) + G_{22}^{\sigma', \Delta'}(x, x) ,
    \end{align}
    which is then used to compute $\Gamma = W + \mu\cdot\rho$.
    This completes the workflow to first order.
    A noteworthy feature of this procedure is that the derivative of $W_0$ is canceled by part of the derivative of $W_1$. 
    This persists to higher orders, with the derivative of $W_{n-1}$ being canceled by part of the derivative of $W_{n}$.
    In particular, the terms with the collective classical fields will yield the cancellations.
    This can be seen diagrammatically by noting that differentiation of any diagram containing a VPT vertex with respect to the collective fields (which, due to their classical nature, act as one-body potentials as far as the graphs are concerned) will yield a diagram with one fewer vertex and an opposite sign from the alternating signs of the cumulant expansion. 

    Because the VPT fields are classical, the double counting challenge from HS fields is avoided.
    However, unlike the classical sources of the inversion method, the classical VPT fields are still collective fields that couple to the fermions in the interacting term, providing an avenue for collective effects to be studied beyond the reach of a standard perturbative expansion.

\section{Implementation details}
We collect here the equations that we numerically solved, with some detail about techniques used.

For inversion at first order, we are simultaneously solving the analogs to the BCS number and gap equations:
        \begin{align}
            \rho &= \int\frac{\dd{k}}{2\pi}\qty(1 - \frac{\xi}{E}) ,\\
            j_0 &= \frac{\LEC}{2}\int\frac{\dd{k}}{2\pi}\frac{j_0}{E}
        .\end{align}
        This is accomplished by minimizing 
        \begin{align}
            \qty(j_0 - \frac{\LEC}{2}\int\frac{\dd{k}}{2\pi}\frac{j_0}{E})^2 + \qty(1 - \frac{1}{\rho}\int\frac{\dd{k}}{2\pi}\qty(1 - \frac{\xi}{E}))^2
        \end{align}
        Minimizing the sum of squares was stable here. 
For inversion at second order, we simultaneously solve 
        \begin{align}
            \rho & = \int\frac{\dd{k}}{2\pi}\qty(1 - \frac{\xi}{E}) ,\\
            j_0 &= \int\frac{\dd{k}}{2\pi}\frac{j_0}{E} + \pdv{\beachball}{\mu_0} \frac{B}{CA - B^2} + \pdv{\beachball}{j_0} \frac{A}{B^2 - CA}
        ,\end{align}
        where $A, B, C$ are defined by Eq.~\eqref{eq:inv_ABC}.
        Both of these minimizations were stable with Nelder-Mead.
        All minimizations here are performed with the Optim package in julia~\cite{mogensen2018optim}.
The second-order contribution to the chemical potential is 
 \begin{align}
     \mu_2 = \pdv{\beachball}{\mu_0}\qty(\frac{C}{AC - B^2}) + \pdv{\beachball}{j_0}\qty(\frac{B}{B^2 - AC})
 ,\end{align}
 and this is added to $\mu_0 + \mu_1 = \mu_0 - \frac{\LEC}{2}\rho$ to get the 2nd order contribution to the chemical potential.

For VPT, at first order we have 
\begin{align}
    \pdv{W_1}{\sigmaVPT} &= -\rho 
    - \frac{\LEC}{2}\rho\pdv{\rho}{\sigmaVPT} 
    - \frac{\LEC}{2}\phi\pdv{\phi}{\sigmaVPT} \notag\\
    &\quad+ \rho 
    + \sigmaVPT\pdv{\rho}{\sigmaVPT} 
    + \DeltaVPT\pdv{\phi}{\sigmaVPT}  \notag\\
    &= 
    %\qty(\sigmaVPT - \frac{\LEC}{2}\rho)
    \mathcal{N}\pdv{\rho}{\sigmaVPT}  
    %\qty(\DeltaVPT - \frac{\LEC}{2}\phi)
    + \mathcal{G}\pdv{\phi}{\sigmaVPT} , \\
    \pdv{W_1}{\DeltaVPT} &= -\phi - \frac{\LEC}{2}\rho\pdv{\rho}{\DeltaVPT} - \frac{\LEC}{2}\phi\pdv{\phi}{\DeltaVPT} + \sigmaVPT\pdv{\rho}{\DeltaVPT} 
        \notag \\ & \qquad \null
    + \phi + \DeltaVPT\pdv{\phi}{\DeltaVPT}  \notag\\
    &= 
    %\qty(\sigmaVPT - \frac{\LEC}{2}\rho)
    \mathcal{N}\pdv{\rho}{\DeltaVPT} 
    %+ \qty(\DeltaVPT - \frac{\LEC}{2}\phi)
    + \mathcal{G}\pdv{\phi}{\DeltaVPT} ,
\end{align}
where $\mathcal{N} \equiv \sigmaVPT - \frac{\LEC\rhot}{2}$ and $\mathcal{G} \equiv \DeltaVPT - \frac{\LEC\phi}{2}$.
These were solved in the same manner as inversion as they are mathematically very similar equations. 
The coefficients to the ``number'' and ``gap'' terms are positive-definite and do not cause any numerical complication.

For VPT, at second order we have
\begin{align}
    \pdv{W_2}{\sigmaVPT} = &\frac{\LEC}{2}\Biggl\{\qty[\qty(\pdv{\rhot}{\sigmaVPT})^2 
    + \qty(\pdv{\phi}{\sigmaVPT})^2]\mathcal{N} \notag\\
    &+ \pdv{\phi}{\sigmaVPT}\qty[\pdv{\rhot}{\sigmaVPT} 
    + \pdv{\phi}{\DeltaVPT}]\mathcal{G}\Biggr\} \notag \\
    &- \frac{1}{2}\pdv[2]{\rhot}{\sigmaVPT}\mathcal{N}^2 
    - \frac{1}{2}\pdv{\phi}{\sigmaVPT}{\DeltaVPT}\mathcal{G}^2 \notag\\
    &- \pdv{\rhot}{\sigmaVPT}{\DeltaVPT}\mathcal{NG} + 
    \pdv{\beachball}{\sigmaVPT} ,\label{eq:dW2_sigma}
\end{align}
and
\begin{align}
    \pdv{W_2}{\DeltaVPT} = &\frac{\LEC}{2}\Biggl\{\pdv{\rhot}{\DeltaVPT}\qty[\qty(\pdv{\phi}{\DeltaVPT}) + \pdv{\rhot}{\sigmaVPT}]\mathcal{N} \notag\\
    + &\qty[\qty(\pdv{\phi}{\DeltaVPT})^2 + \qty(\pdv{\rhot}{\DeltaVPT})^2]\mathcal{G}\Biggr\} \notag \\
    - &\frac{1}{2}\pdv{\rhot}{\DeltaVPT}{\sigmaVPT}\mathcal{N}^2 
    - \frac{1}{2}\pdv[2]{\phi}{\DeltaVPT}\mathcal{G}^2 \notag\\
    - &\pdv{\phi}{\DeltaVPT}{\sigmaVPT}\mathcal{NG} + \pdv{\beachball}{\DeltaVPT} ,\label{eq:dW2_Delta}
\end{align}
where $\pdv*{\rhot}{\DeltaVPT} = \pdv*{\phi}{\sigmaVPT}$, so they can be used interchangeably. 
The beachball diagram is basically the same form as we had for inversion in Eq.~\eqref{eq:bb}, the only difference is $\xi$ is now defined as $\frac{k^2}{2m} - \mu - \sigmaVPT$.

A stable way to solve for when Eqs.~\eqref{eq:dW2_sigma} and \eqref{eq:dW2_Delta} equal zero is to treat them as components of the gradient of $W$ as a function of $\sigmaVPT, \DeltaVPT$ at given input $\mu$ values, and then minimizing the approximation to the full $W$ with respect to the VPT parameters.
So to find the second order VPT parameters, we minimized 
\begin{align}
    W & \approx \int_k(\xi - E) 
    - \frac{\LEC}{4}(\rhot^2 + \phi^2) 
    + \sigmaVPT\rhot + \DeltaVPT\phi \notag \\
    &\qquad\null - \frac{1}{2}\pdv{\rhot}{\sigmaVPT}\mathcal{N}^2\notag \\
    &\qquad\null - \frac{1}{2}\pdv{\phi}{\DeltaVPT}\mathcal{G}^2 \notag \\
    &\qquad\null - M\mathcal{N}\mathcal{G} + \beachball,
\end{align}
where, in terms of the input parameters, 
\begin{align}
    \rhot &= \int\frac{\dd{k}}{2\pi}\qty(1 - \frac{\frac{k^2}{2m} - \mu - \sigmaVPT}{\sqrt{\qty(\frac{k^2}{2m} - \mu - \sigmaVPT)^2 + \DeltaVPT^2}})\\
    \phi &= \int\frac{\dd{k}}{2\pi}\frac{\DeltaVPT}{\sqrt{\qty(\frac{k^2}{2m} - \mu - \sigmaVPT)^2 + \DeltaVPT^2}}
.\end{align}
The optimiziation was performed in two steps: 
\begin{enumerate}
    \item An initial, softer optimization using Adam with the first-order values of $\sigmaVPT,\DeltaVPT$ at a given chemical potential as an initial guess. The learning rate for Adam was 0.001.
    \item A second, more strict optimization using BFGS with the output from Adam as the initial guess.
\end{enumerate}
Criteria for termination of the algorithms were smallness of the gradient and the size of steps taken. 
Adaptive quadrature was used to accommodate the changing of the curves for the diagrams.
For the beachball, a Genz-Malik rule was implemented through the \texttt{Cubature.jl} package~\cite{cubature}.

\bibliography{BRST_refs_2019}
 
\end{document}